\documentclass[11pt,a4paper]{article}

\usepackage[T1]{fontenc}
\usepackage[utf8]{inputenc}
\usepackage[a4paper,margin=2.5cm]{geometry}
\usepackage{graphicx}
\usepackage{amsmath,amssymb}
\usepackage{newtxtext}
\usepackage{newtxmath}
\usepackage{bm}
\usepackage{dcolumn}
\usepackage{enumitem}
\usepackage{authblk}
\usepackage{xcolor}
\usepackage{cite}
\usepackage{hyperref}

\hypersetup{
    colorlinks = true,
    linkcolor  = black,
    citecolor  = black,
    urlcolor   = blue,
}

\begin{document}

\title{Adaptive near-contact repulsion in conservative Allen--Cahn phase-field lattice Boltzmann multiphase model}

% Author and affiliation formatting
\renewcommand{\Authfont}{\normalsize\bfseries}
\renewcommand{\Affilfont}{\normalfont\small}
\setlength{\affilsep}{0.45em}

\author[1]{Andrea Montessori}
\author[2]{Maria Rosa Lisboa}
\author[3]{Marco Lauricella}
\author[4,5]{Sauro Succi}

\affil[1]{Department of Civil, Computer Science and Aeronautical Technologies Engineering, Roma Tre University, Via Vito Volterra, 00146 Rome, Italy}
\affil[2]{Pontifícia Universidade Católica do Paraná, Rua Imaculada Conceição, 1155, Curitiba 80215-901, Brazil}
\affil[3]{Istituto per le Applicazioni del Calcolo, Consiglio Nazionale delle Ricerche, Via dei Taurini 19, 00185 Rome, Italy}
\affil[4]{Center for Life Nano- \& Neuro-Science, Fondazione Istituto Italiano di Tecnologia, Viale Regina Elena 291, 00161 Rome, Italy}
\affil[5]{PoreLab, Department of Physics, Norwegian University of Science and Technology, 7491 Trondheim, Norway}

\date{}
\maketitle
\vspace{-1.4em}
\begin{center}
\small\textit{Corresponding author:} \href{mailto:andrea.montessori@uniroma3.it}{andrea.montessori@uniroma3.it}
\end{center}
\vspace{0.5em}

\begin{abstract}
Unresolved thin-film dynamics often causes spurious coalescence in diffuse-interface simulations of multiphase flows.
We address this issue by introducing a fully local repulsive near-contact flux in a conservative Allen--Cahn phase-field model coupled to lattice Boltzmann hydrodynamics.
The interaction activates only for nearby interfaces with opposite orientations, with a strength that self-adjusts according to a closed-form phase-field-based indicator of local interface proximity.
The resulting method avoids nonlocal geometric procedures, retains computational efficiency, and is well suited to massively parallel implementations.
Tests on collision benchmarks and three-dimensional bubble swarms demonstrate robust suppression of artificial merging and physically consistent near-contact dynamics.
\end{abstract}

\noindent\textbf{Keywords:} lattice Boltzmann method; conservative Allen--Cahn equation; phase-field method; near-contact interaction; multiphase flows; droplet collisions.

%\linenumbers

\section{Introduction}

Multiphase flows are central to a broad range of natural phenomena and environmental, engineering, and life-science applications. 
They play a key role in cloud physics \cite{shaw2003particle}, aerosol transport \cite{guha2008transport}, boiling and condensation \cite{kharangate2017review}, emulsification \cite{bogdan2022stochastic,haakansson2019emulsion}, bubble-driven mixing \cite{risso2018agitation}, 
microfluidics \cite{sattari2020multiphase}, porous-media transport \cite{miller1998multiphase}, and energy-related technologies such as carbon sequestration and subsurface storage \cite{ringrose2021storage}. 
In all these settings, the dynamics of interfaces controls momentum exchange, mass transport, topology changes, and ultimately the large-scale behavior of the system. 
A predictive description of multiphase flows is, therefore, of major importance both for advancing fundamental understanding and for enabling the rational design of technological processes.

However, numerical investigation of such flows remains highly challenging \cite{tiribocchi2025lattice, balachandar2010turbulent,subramaniam2020multiphase}. 
The main difficulty lies in the need to accurately capture the interplay between physical mechanisms of different nature that act on widely separated spatial and temporal scales \cite{subramaniam2020multiphase}. 
At the macroscopic level, inertia, viscous stresses, capillary forces, and buoyancy govern the global flow evolution. 
At the same time, many crucial events are controlled by localized interfacial processes occurring over much smaller scales, such as thin-film drainage, interface deformation, rupture, and near-contact interactions \cite{craster2009dynamics,dai2008disjoining,boinovich2002forces,tirumkudulu2004role}. 
These processes often determine whether two interfaces merge, rebound, slide past each other, or remain separated, thereby strongly affecting the overall evolution of the system. 
The resulting cross-scale coupling poses a major challenge for numerical models, especially in three-dimensional simulations and in the presence of large density and viscosity contrasts.

These considerations motivate the development of accurate, high-fidelity, and computationally efficient numerical methods capable of resolving interface dynamics while also accounting for physical effects that cannot be directly represented within the standard continuum description based on the Navier--Stokes equations alone \cite{tiribocchi2025lattice}. 
Although classical hydrodynamic models provide the correct large-scale framework, they are not sufficient, by themselves, to describe unresolved interfacial physics arising at scales below the numerical grid resolution, particularly during near-contact events. 
As a result, there is a clear need for computational approaches that combine robust mesoscale hydrodynamic solvers, accurate interface-capturing formulations, and suitable closures for subgrid interfacial phenomena. 
Moreover, such approaches must also retain a high degree of locality and algorithmic efficiency to remain viable for large-scale simulations on modern high-performance computing architectures \cite{succi2019towards}.

The aim of the present paper is to develop such a framework in the context of diffuse-interface multiphase flow modeling. 
More specifically, we introduce a repulsive near-contact flux \cite{montessori2019mesoscale} within a conservative Allen--Cahn phase-field formulation \cite{jeong2017conservative,liang2018phase,liang2019lattice} coupled to a lattice Boltzmann solver for the hydrodynamics \cite{benzi1992lattice,kruger2017lattice}. 
The proposed model is designed to prevent spurious coalescence when two interfaces approach each other across an unresolved thin film, while preserving the locality, robustness, and computational efficiency of the underlying numerical method. 
The central idea is to account for unresolved near-contact physics through a short-range repulsive contribution that is activated only whenever two diffuse interfaces come sufficiently close and exhibit opposite orientation, namely in configurations compatible with the formation of an intervening liquid film. 
In this way, the model selectively targets those interfacial interactions for which standard diffuse-interface descriptions are most prone to artificial merging, while remaining inactive elsewhere and therefore minimally intrusive on the resolved hydrodynamics.

A distinctive feature of the proposed formulation is that local interface proximity is characterized without requiring explicit geometric reconstruction, ray tracing along interface normals, or other nonlocal procedures. Instead, an effective film-thickness indicator is obtained in closed form from the local phase-field profile, thereby preserving the compact stencil and locality of the numerical scheme. This construction also endows the repulsive interaction with a self-tuning character: the magnitude of the near-contact flux is not prescribed as a fixed ad hoc correction, but adapts dynamically to the local degree of interface overlap and to the evolving near-contact configuration.
As the drainage-driven approach between interfaces becomes stronger and the effective film thickness decreases, the repulsive contribution automatically intensifies, counteracting further collapse of the intervening layer; conversely, when the interfaces separate, the flux rapidly vanishes. 
The resulting mechanism provides a coarse-grained but physically consistent closure of unresolved thin-film stabilization, capable of balancing the spurious attractive tendency induced by insufficient spatial resolution without introducing long-range artifacts or unnecessary distortion of the bulk flow.

Because the entire formulation is local, explicit, and free of costly geometric operations, it is particularly well suited for thread-safe and massively parallel implementations \cite{montessori2024high,lauricella2025thread}. 
This makes it especially attractive for large-scale three-dimensional simulations of complex interfacial flows, where many-body near-contact events occur frequently and where numerical efficiency is as critical as physical fidelity.

The paper is organized as follows. 
Section Methods presents the numerical methodology, including the hydrodynamic lattice Boltzmann solver, the conservative Allen--Cahn phase-field formulation, and the derivation and implementation of the repulsive near-contact flux. 
Section Results assesses the performance of the model through representative benchmark problems and application cases, highlighting its ability to capture collision dynamics and many-body interfacial interactions without spurious merging. 
Finally, in the Conclusions we summarize the main findings and discuss perspectives for future developments.
%\nocite{*}

\section{Methods}

\subsection{Coupled recursive-regularized thread-safe lattice Boltzmann and finite-difference MUSCL scheme for the Navier--Stokes and conservative Allen--Cahn equations} \label{sec:tslb}

We solve the hydrodynamics of the two-component system through a thread-safe lattice Boltzmann (LB) method,
implemented in \textit{accLB} \cite{lauricella2025acclb,montessori2026breakdown} and coupled to an interface-capturing phase-field solver (described later).
The LB solver is designed to (i) remain fully local on shared-memory architectures such as GPUs, thereby avoiding
memory race conditions during streaming/collision, and (ii) improve stability at low viscosities through a third-order recursive regularization of non-equilibrium moments \cite{montessori2024high}.

\textit{Hydrodynamic limit.}
In the low-Mach, small-Knudsen limit, the discrete kinetic dynamics recovers the variable-density,
variable-viscosity Navier--Stokes equations \cite{fakhari2017improved,zu2013phase}
\begin{align}
\partial_t \rho + \nabla\cdot(\rho \mathbf{u}) &= 0, \label{eq:conti}\\
\partial_t(\rho \mathbf{u}) + \nabla\cdot(\rho \mathbf{u}\otimes \mathbf{u}) &=
-\nabla p + \nabla\cdot\!\left[\rho \nu\left(\nabla\mathbf{u}+(\nabla\mathbf{u})^{T}\right)\right]
+\mathbf{F}_{\sigma} + \mathbf{F}_{\rm rep} + \mathbf{F}_{\rm ext}. \label{eq:mom}
\end{align}
Here $\rho$ is the mixture density, $\mathbf{u}$ the velocity, $\nu$ the kinematic viscosity,
and $\mathbf{F}_{\sigma}$ is the capillary (surface-tension) force density. The term $\mathbf{F}_{\rm rep}$
denotes the near-contact repulsive contribution introduced in Sec.~\ref{sec:nci_theory}, while $\mathbf{F}_{\rm ext}$
collects possible external body forcings.

\textit{Discrete kinetic equation and thread-safe update.}
We consider a $q$-speed stencil (D3Q27) with discrete velocities $\{\mathbf{c}_i\}_{i=0}^{q-1}$,
weights $\{w_i\}$, and lattice time step $\Delta t=1$.
The LB dynamics reads \cite{kruger2017lattice,benzi1992lattice}
\begin{equation}
f_i(\mathbf{x}+\mathbf{c}_i, t+1) =
f_i(\mathbf{x}, t) + \omega\!\left[f_i^{\rm eq}(\mathbf{x},t)-f_i(\mathbf{x},t)\right] + (1-\frac{\omega}{2})S_i(\mathbf{x},t),
\label{eq:lbe}
\end{equation}
where $f_i$ are the populations, $\omega$ is the relaxation frequency, and $S_i$ is a forcing source term \cite{guo2002discrete}. 
The kinematic viscosity is linked to $\omega$ through
\begin{equation}
\nu = c_s^2\left(\frac{1}{\omega}-\frac{1}{2}\right),
\label{eq:nuomega}
\end{equation}
with $c_s$ the lattice sound speed \cite{succi2018lattice}.
Following the thread-safe paradigm, both the equilibrium and non-equilibrium parts are reconstructed
from local macroscopic fields via a Hermite expansion (see below), so that each lattice node can be
updated independently without reading/writing neighboring populations (single-pass streaming--collision,
no flip-flop buffering). This eliminates race conditions on shared-memory parallel architectures.

\textit{Macroscopic fields.}
We employ the incompressible-pressure formulation \cite{zu2013phase} where the kinetic zeroth moment yields a
dimensionless pressure-like field $p^{*}$, while the physical pressure is
\begin{equation}
p = \rho\, c_s^2\, p^{*}.
\label{eq:pstar}
\end{equation}
The macroscopic fields are recovered as
\begin{align}
p^{*}(\mathbf{x},t) &= \sum_i f_i(\mathbf{x},t), \label{eq:pstar_moment}\\
\mathbf{u}(\mathbf{x},t) &= \sum_i f_i(\mathbf{x},t)\,\mathbf{c}_i + \frac{1}{2}\sum_i S_i(\mathbf{x},t)\,\mathbf{c}_i.
\label{eq:u_moment}
\end{align}
The density $\rho$ and viscosity $\nu$ are space-dependent and determined from the phase field (mixture law),
which will be detailed together with the interface-transport equation in the next section.

\textcolor{black}{An additional advantage of the present velocity-based incompressible formulation is its weak sensitivity to density and viscosity contrasts in terms of equilibrium spurious currents. Since velocity is kinetically evolved, whereas density is prescribed independently through the phase field and pressure is represented by a separate pressure-like moment, these fields remain largely decoupled at the kinetic level. For the configurations considered here, equilibrium spurious velocities remain of the order of $10^{-6}$ in lattice units, with no appreciable variation over the investigated density and viscosity ratios. More generally, the magnitude and spatial extent of parasitic interfacial currents depend on the specific multiphase formulation and on the isotropy properties of its discrete operators. In pseudopotential models, for example, the isotropy of the interaction force and of the associated lattice pressure tensor plays a central role. Lulli et al.~\cite{lulli2021structure} showed that an appropriate construction of the pressure tensor can substantially reduce both the intensity and spatial extent of spurious currents. Such pressure-tensor-based strategies address equilibrium lattice artifacts and are therefore complementary to the present near-contact treatment, which instead targets premature coalescence caused by unresolved thin-film dynamics.}

\textit{Mixture properties and relaxation frequency.}
The local material properties are functions of the phase field $\phi\in[0,1]$ (with $\phi=0$ in the gas
and $\phi=1$ in the liquid). In the present work we interpolate the \emph{dynamic} viscosity linearly as \cite{fakhari2017improved}
\begin{equation}
\mu(\phi)=\mu_g + \left(\mu_\ell-\mu_g\right)\phi .
\label{eq:mu_mix_linear}
\end{equation}
The kinematic viscosity then follows from $\nu(\phi)=\mu(\phi)/\rho(\phi)$, i.e.
\begin{equation}
\nu(\phi)=\frac{\mu_g + (\mu_\ell-\mu_g)\phi}{\rho(\phi)}.
\label{eq:nu_from_mu}
\end{equation}
Accordingly, the relaxation frequency is space-dependent and given by
\begin{equation}
\omega(\phi)=\left(\frac{1}{2}+\frac{\nu(\phi)}{c_s^2}\right)^{-1},
\qquad
\tau(\phi)=\omega(\phi)^{-1}.
\label{eq:omega_phi}
\end{equation}
The density field $\rho(\phi)$ is obtained from the phase-field mixture rule (reported together with the
phase-field model in the next section), and is used consistently in the pressure relation $p=\rho c_s^2 p^{*}$.

\textit{Third-order equilibrium.}
The equilibria are obtained from a third-order (Mach-number) Hermite expansion \cite{montessori2024high,montessori2015lattice,malaspinas2015increasing}:
\begin{equation}
f_i^{\rm eq} = w_i\Bigg[
p^{*}
+\frac{\mathbf{c}_i\!\cdot\!\mathbf{u}}{c_s^2}
+\frac{\mathcal{H}_i^{(2)}:\left(\mathbf{u}\otimes\mathbf{u}\right)}{2c_s^4}
+\frac{\mathcal{H}_i^{(3)}\;\vdots\;\left(\mathbf{u}\otimes\mathbf{u}\otimes\mathbf{u}\right)}{6c_s^6}
\Bigg],
\label{eq:feq3}
\end{equation}
where ''$:$'' denotes double contraction between second-order tensors and ''$\vdots$'' the triple contraction
between third-order tensors. The second- and third-order Hermite tensors are
\begin{align}
\mathcal{H}_i^{(2)} &= \mathbf{c}_i\otimes\mathbf{c}_i - c_s^2 \mathbf{I}, \label{eq:H2}\\
\mathcal{H}_i^{(3)} &= \mathbf{c}_i\otimes\mathbf{c}_i\otimes\mathbf{c}_i - c_s^2\,\mathcal{S}(\mathbf{c}_i\otimes\mathbf{I}),
\label{eq:H3}
\end{align}
with $\mathbf{I}$ the identity tensor and $\mathcal{S}(\cdot)$ the full symmetrization over index permutations
(e.g.\ $\mathcal{S}(\mathbf{c}\otimes\mathbf{I})=\mathbf{c}\otimes\mathbf{I}+\mathbf{I}\otimes\mathbf{c}+\cdots$).

\textit{Recursive regularization of non-equilibrium populations.}
We decompose $f_i=f_i^{\rm eq}+f_i^{\rm neq}$ and reconstruct $f_i^{\rm neq}$ up to third order as
\begin{equation}
f_i^{\rm neq} = w_i\Bigg[
\frac{\mathcal{H}_i^{(2)}:\mathbf{A}^{(2)}_{\rm neq}}{2c_s^4}
+\frac{\mathcal{H}_i^{(3)}\;\vdots\;\mathbf{A}^{(3)}_{\rm neq}}{6c_s^6}
\Bigg],
\label{eq:fneq3}
\end{equation}
where $\mathbf{A}^{(2)}_{\rm neq}$ is the (symmetric) second-order non-equilibrium moment and
$\mathbf{A}^{(3)}_{\rm neq}$ is obtained via Hermite recursivity $\sum_i \mathcal{H}_i^{(2)}f_i^{neq}$, where $f_i^{neq}=f_i-f_i^{eq} + 0.5S_i$ \cite{szeg1939orthogonal,feng2019pre}.
At the Navier--Stokes level, $\mathbf{A}^{(2)}_{\rm neq}$ can be expressed in gradient form as
\begin{equation}
\mathbf{A}^{(2)}_{\rm neq} = -\frac{c_s^2}{\omega}\left[\nabla(\mathbf{u})+\nabla(\mathbf{u})^{T}\right],
\label{eq:A2grad}
\end{equation}
while the third-order tensor is obtained recursively as
\begin{equation}
\mathbf{A}^{(3)}_{\rm neq} = \mathcal{S}\!\left(\mathbf{u}\otimes \mathbf{A}^{(2)}_{\rm neq}\right).
\label{eq:A3rec}
\end{equation}
In practice, the regularization acts as a projection of $f_i^{\rm neq}$ onto the Hermite subspace
spanned by moments up to third order, filtering out higher-order ghost modes and improving robustness
at low viscosity/high Reynolds number regimes.

\textit{Forcing scheme and multiphase force decomposition.}
Forcing is incorporated via the Guo source term \cite{guo2002discrete},
\begin{equation}
S_i = w_i\left[
\frac{\mathbf{c}_i-\mathbf{u}}{c_s^2} + \frac{(\mathbf{c}_i\!\cdot\!\mathbf{u})}{c_s^4}\,\mathbf{c}_i
\right]\cdot \frac{\mathbf{F}}{\rho},
\label{eq:guo}
\end{equation}
and implemented consistently with the trapezoidal rule by shifting the equilibria.

For multiphase flows with strong density/viscosity inhomogeneities, we decompose \cite{zu2013phase}
\begin{equation}
\mathbf{F} = \mathbf{F}_{\sigma} + \mathbf{F}_{p} + \mathbf{F}_{\nu} + \mathbf{F}_{\rm rep} + \mathbf{F}_{\rm ext}.
\label{eq:Fsplit}
\end{equation}
The capillary contribution is expressed in chemical-potential form \cite{jacqmin1999calculation},
\begin{equation}
\mathbf{F}_{\sigma} = \mu_{\phi}\,\nabla \phi,
\label{eq:Fcap}
\end{equation}
where $\phi$ is the phase field and $\mu_{\phi}$ its chemical potential (defined together with the phase-field model).

Because the pressure is given by $p=\rho c_s^2 p^{*}$, its gradient splits as
$\nabla p = \rho c_s^2 \nabla p^{*} + p^{*}c_s^2 \nabla \rho$.
The first term is embedded in the choice of equilibria \eqref{eq:feq3}, whereas the second term must be included
explicitly as a correction force
\begin{equation}
\mathbf{F}_{p} = -p^{*}c_s^2\,\nabla \rho.
\label{eq:Fp}
\end{equation}
Finally, to reproduce the variable-coefficient viscous operator in \eqref{eq:mom},
an additional correction proportional to $\nabla\rho$ is introduced by exploiting the link between
the second-order kinetic moment and the deviatoric stress \cite{kruger2009shear}. In compact tensor form, this reads:
\begin{equation}
\mathbf{F}_{\nu} =
-\frac{\nu\omega}{c_s^2}\left[\sum_i \left(f_i-f_i^{\rm eq}\right)\left(\mathbf{c}_i\otimes\mathbf{c}_i\right)\right]\cdot\nabla\rho,
\label{eq:Fnu}
\end{equation}
where ''$\cdot$'' denotes the contraction of the second-order tensor with the vector $\nabla\rho$.
The near-contact force $\mathbf{F}_{\rm rep}$ is discussed and modeled in Sec.~\ref{sec:nci_theory}.

\textit{Discrete derivatives.}
Whenever gradients/Laplacians of scalar fields (e.g.\ $\rho$, $\phi$, $\mu_{\phi}$) are required in the forcing terms,
we employ isotropic lattice-consistent stencils \cite{shan2006analysis, philippi2006from, thampi2013isotropic}: 
\begin{align}
\nabla \Psi(\mathbf{x}) &= \frac{1}{c_s^2}\sum_i w_i\,\Psi(\mathbf{x}+\mathbf{c}_i)\,\mathbf{c}_i,
\label{eq:grad_iso}\\
\nabla^2 \Psi(\mathbf{x}) &= \frac{2}{c_s^2}\left(\sum_{i\neq 0} w_i\,\Psi(\mathbf{x}+\mathbf{c}_i) - (1-w_0)\,\Psi(\mathbf{x})\right),
\label{eq:lap_iso}
\end{align}
for a generic scalar field $\Psi$.

\subsubsection{MUSCL scheme discretization for the conservative Allen--Cahn equation}

\textit{Conservative Allen--Cahn equation.}
The interface dynamics is captured through a phase field $\phi\in[0,1]$, with $\phi=1$ identifying the liquid phase and $\phi=0$ the gas phase; the diffuse interface is conventionally located at $\phi_0=1/2$.
\textcolor{black}{The conservative Allen--Cahn equation can be advanced either through a dedicated lattice Boltzmann equation or by means of conventional spatial discretizations. Phase-field-based kinetic formulations have been successfully developed, including the multiple-relaxation-time lattice Boltzmann model proposed by Liang et al.~\cite{liang2014phasefield} for incompressible multiphase flows. In the present framework, the hydrodynamics is solved using the thread-safe recursive-regularized lattice Boltzmann method, whereas the phase-field equation is advanced through a finite-difference MUSCL--TVD scheme. This hybrid choice provides conservative and robust phase-field advection while retaining a compact stencil and a highly parallel implementation.}

Following our previous works \cite{lauricella2025thread}, the evolution of $\phi$ is governed by the \emph{conservative Allen--Cahn equation} (CACE) \cite{chiu2011conservative}:
\begin{equation}
\frac{\partial \phi}{\partial t}
+\mathbf{u}\cdot\nabla \phi
=
D\,\nabla^2\phi
-\kappa\,\nabla\cdot\!\Big(\phi(1-\phi)\,\mathbf{n}\Big),
\label{eq:cace}
\end{equation}
where $D$ is a (constant) interface diffusivity and $\mathbf{n}$ is the interface normal,
\begin{equation}
\mathbf{n}=\frac{\nabla\phi}{|\nabla\phi|}.
\label{eq:normal}
\end{equation}
The last term on the right-hand side of \eqref{eq:cace} is a compressive (anti-diffusive) flux that counteracts numerical diffusion and maintains a sharp, yet smooth, hyperbolic-tangent-like interface profile across a prescribed thickness $\delta$.
A common choice, adopted here, links the compression strength to the diffusivity via
\begin{equation}
\kappa=\frac{4D}{\delta},
\label{eq:kappa_delta}
\end{equation}
so that the interface remains close to the target thickness throughout the simulation.

\textit{Conservative flux form.}
For the sake of numerical robustness, we cast the advective contribution in conservative form. Introducing the advective flux $\mathbf{F}_a=\mathbf{u}\phi$, Eq.~\eqref{eq:cace} can be rewritten as
\begin{equation}
\frac{\partial \phi}{\partial t}+\nabla\cdot(\mathbf{u}\phi)
=
D\,\nabla^2\phi
-\kappa\,\nabla\cdot\!\Big(\phi(1-\phi)\,\mathbf{n}\Big),
\label{eq:cace_conservative}
\end{equation}
so that mass conservation is enforced at the discrete level by a flux-difference update.

\textit{Discrete update in lattice units.}
In the implementation, the update at a fluid node $(i,j,k)$ takes the form
\begin{equation}
\phi^{n+1}_{ijk}
=
\phi^{n}_{ijk}
-
(\nabla_h\!\cdot \mathbf{F}_a)_{ijk}
+
\tau_{\mathrm{diff}}\,(\nabla_h^2 \phi)_{ijk}
+
S^{\mathrm{comp}}_{ijk},
\label{eq:phi_update_muscl}
\end{equation}
where $\nabla_h$ denotes discrete operators on the Cartesian lattice, $\tau_{\mathrm{diff}}$ collects the constants multiplying the discrete Laplacian (consistent with $D$ in \eqref{eq:cace}), and $S^{\mathrm{comp}}$ represents the discrete counterpart of the compressive term $-\kappa\nabla\cdot(\phi(1-\phi)\mathbf{n})$.
A double-buffering strategy is adopted, $\phi(\texttt{flip})\to\phi(\texttt{flop})$, and the update is applied only to fluid nodes (solid nodes are skipped).

\subsubsection*{3D MUSCL--TVD (minmod) conservative advection for $\phi$}

\textit{Face fluxes and discrete divergence.}
Assuming $\Delta x=\Delta y=\Delta z=1$, the discrete divergence of the advection flux is computed as
\begin{equation}
(\nabla_h\!\cdot\mathbf{F}_a)_{ijk}
=
\left(F^x_{i+\frac12,j,k}-F^x_{i-\frac12,j,k}\right)
+
\left(F^y_{i,j+\frac12,k}-F^y_{i,j-\frac12,k}\right)
+
\left(F^z_{i,j,k+\frac12}-F^z_{i,j,k-\frac12}\right),
\label{eq:divF}
\end{equation}
with face fluxes
\begin{equation}
F^x_{i+\frac12,j,k}=u_{i+\frac12,j,k}\,\phi^{\mathrm{up}}_{i+\frac12,j,k},\qquad
F^y_{i,j+\frac12,k}=v_{i,j+\frac12,k}\,\phi^{\mathrm{up}}_{i,j+\frac12,k},\qquad
F^z_{i,j,k+\frac12}=w_{i,j,k+\frac12}\,\phi^{\mathrm{up}}_{i,j,k+\frac12}.
\label{eq:face_flux_def}
\end{equation}
The face-centered velocities are obtained by arithmetic averages of cell-centered values, e.g.
$u_{i+\frac12,j,k}=\tfrac12(u_{i,j,k}+u_{i+1,j,k})$ (and similarly for $v$ and $w$).

\textit{MUSCL reconstruction with minmod limiter.}
The upwind face value $\phi^{\mathrm{up}}$ is computed with a MUSCL piecewise-linear reconstruction, limited by minmod to guarantee the TVD property.
Along the $x$ direction, the limited slope at cell $(i,j,k)$ reads as:
\begin{equation}
s^x_{i,j,k}=\mathrm{minmod}\!\left(\phi_{i,j,k}-\phi_{i-1,j,k},\;\phi_{i+1,j,k}-\phi_{i,j,k}\right),
\label{eq:slope_minmod_x}
\end{equation}
where
\begin{equation}
\mathrm{minmod}(a,b)=
\begin{cases}
0, & ab\le 0,\\
\mathrm{sign}(a)\,\min(|a|,|b|), & ab>0.
\end{cases}
\label{eq:minmod}
\end{equation}
The left/right reconstructed states at the face $i+\tfrac12$ are then
\begin{equation}
\phi^{L}_{i+\frac12,j,k}=\phi_{i,j,k}+\tfrac12 s^x_{i,j,k},
\qquad
\phi^{R}_{i+\frac12,j,k}=\phi_{i+1,j,k}-\tfrac12 s^x_{i+1,j,k},
\label{eq:recon_lr_x}
\end{equation}
and the upwind selection is
\begin{equation}
\phi^{\mathrm{up}}_{i+\frac12,j,k}=
\begin{cases}
\phi^{L}_{i+\frac12,j,k}, & u_{i+\frac12,j,k}\ge 0,\\
\phi^{R}_{i+\frac12,j,k}, & u_{i+\frac12,j,k}< 0.
\end{cases}
\label{eq:upwind_select_x}
\end{equation}
The same procedure is applied along $y$ and $z$ by replacing $(i,u)$ with $(j,v)$ and $(k,w)$, respectively.
This scheme is second-order accurate in smooth regions, reverts to first order near sharp gradients, and prevents spurious oscillations at the interface.

\textit{Stencil requirements and boundary handling.}
The minmod slopes at $i\pm1$ require accessing $\phi$ up to $i\pm2$ (and analogously in $y$ and $z$); hence, two buffer layers are required in each direction.

In periodic domains, buffer layers are filled by wrap-around; otherwise, they are filled consistently with the imposed boundary conditions.

\subsubsection*{Discrete compressive term and isotropic divergence}

The compressive contribution in \eqref{eq:cace} can be interpreted as the divergence of a \emph{compressive flux}
$\mathbf{F}_c=\phi(1-\phi)\mathbf{n}$.
In the code, its discrete divergence is evaluated with an isotropic D3Q27-based stencil, combining axis-, face-diagonal-, and body-diagonal contributions with weights $(p_1,p_2,p_3)$.
Denoting by $\mathbf{a}=(a_x,a_y,a_z)$ the discrete representation of $\mathbf{F}_c$ (or an equivalent compressive vector field), we compute
\begin{equation}
S^{\mathrm{comp}}_{ijk}
=
-\frac{C_{sharp}}{c_s^2}\,(\nabla_h\!\cdot\mathbf{a})_{ijk},
\label{eq:source_sharp}
\end{equation}
where $C_{sharp}$ is the sharpening coefficient and $c_s$ the lattice sound speed.
The isotropic stencil reduces grid-orientation artifacts and yields a compressive contribution consistent with the discrete operators used by the LB solver.

\subsection{Near-contact interaction model: theory and continuum formulation}
\label{sec:nci_theory}

\subsubsection{Physical motivation: unresolved thin-film stabilization as a coarse-grained normal stress}
\label{sec:phys_motiv}

In three-dimensional multiphase flows involving collisions and near-contact events (droplet impacts, dense emulsions, bubbly flows),
the transition between coalescence and rebound is controlled by the dynamics of the thin intervening film that forms between two
approaching interfaces \cite{QianLaw1997, orme1997experiments}. Coalescence requires this film to drain down to a critical thickness at which rupture is triggered by
short-range intermolecular attraction (and/or by instability mechanisms), whereas rebound becomes possible if the film remains stable
long enough for the interfaces to re-separate. Importantly, such a transition is not restricted to surfactant-laden systems: even in
''clean'' binary-fluid collisions, variations of fluid properties and droplet size can shift the outcome because they modify the
characteristic drainage time and the attainable minimum film thickness.

Diffuse-interface simulations cannot directly resolve the nanometric scales associated with thin-film drainage and rupture \cite{jacqmin1999calculation, craster2009dynamics}. As a consequence,
interfaces may merge prematurely for purely numerical reasons (insufficient resolution of the film, excessive overlap of diffuse interfaces, or missing unresolved repulsive physics), yielding spurious coalescence in regimes where physical systems would bounce \cite{li2016macroscopic}.
To represent the stabilizing action of unresolved short-range interactions (steric/electrostatic/disjoining effects) and, more broadly, to \emph{control} coalescence by delaying film collapse, we introduce, at the mesoscopic level, a near-contact interaction (NCI) in the form of an additional \emph{purely normal} body force $\bm{F}_{\mathrm{rep}}$. Such a force activates only when two diffuse interfaces approach each other with opposite orientations.

\subsubsection{Interfacial indicator and geometric ingredients}
\label{sec:indicator_geometry}

We define the standard interfacial indicator
\begin{equation}
  q(\phi)=\phi(1-\phi)\in[0,1/4],
  \label{eq:q_def}
\end{equation}
which is maximal at $\phi=1/2$ and vanishes in the bulk.

The key modeling requirement is to distinguish (i) an isolated interface (where $q$ is nonzero but no thin film exists) from
(ii) a true near-contact film configuration (where two interfaces face each other with opposite orientation). To this end, we introduce:

\begin{itemize}
  \item an \emph{activation function} $\chi_{\mathrm{nc}}(\bm{x})\in\{0,1\}$ that identifies near-contact regions;
  \item a \emph{partner mapping} $\bm{x}_p(\bm{x})$ that associates to an activated point $\bm{x}$ a nearby point $\bm{y}=\bm{x}_p(\bm{x})$
        located on the opposing interface across the film.
\end{itemize}

At the continuum level, $\chi_{\mathrm{nc}}$ may be understood as indicating the existence, within a local neighborhood of $\bm{x}$,
of a second interfacial location $\bm{y}$ such that $q(\bm{y})$ is comparable to $q(\bm{x})$ and $\bm{n}(\bm{y})$ is sufficiently
opposite to $\bm{n}(\bm{x})$. In practice, we realize this concept through a bounded local search on the lattice (no ray tracing);
the discrete construction is reported later, but the theory below is expressed directly in terms of $\chi_{\mathrm{nc}}$ and
$\bm{x}_p(\bm{x})$.

\subsubsection{Definition of the repulsive force: magnitude, facing factor, and antisymmetric normal direction}
\label{sec:rep_force_def}

Let $\bm{y}=\bm{x}_p(\bm{x})$ be the partner point associated with $\bm{x}$ when $\chi_{\mathrm{nc}}(\bm{x})=1$.
We define the paired interfacial indicator
\begin{equation}
  q_1 = q(\phi(\bm{x})),\qquad q_2 = q(\phi(\bm{y})),\qquad
  q_{\mathrm{pair}}(\bm{x}) = \frac{1}{2}\bigl(q_1+q_2\bigr).
  \label{eq:q_pair}
\end{equation}
Using $q_{\mathrm{pair}}$ (rather than $q_1$ alone) improves symmetry across the film and reduces sensitivity to small
discretization asymmetries.

\textit{Facing factor (activation strength).}
A scalar facing factor is introduced to modulate the interaction according to the degree of opposition of the two interfaces:
\begin{equation}
  f_{\mathrm{face}}(\bm{x})
  =
  \max\!\Bigl(0,\,-\bm{n}(\bm{x})\cdot\bm{n}(\bm{y})\Bigr)\in[0,1].
  \label{eq:facing_factor}
\end{equation}
This factor is zero when normals are not opposing and approaches unity for nearly antiparallel normals.

\textit{Symmetric normal direction (antisymmetry and purely normal forcing).}
To ensure that the repulsion acts predominantly along the film-normal direction and exhibits an action--reaction structure, we define the
symmetric direction
\begin{equation}
  \bm{n}_{\mathrm{sym}}(\bm{x})
  =
  \frac{\bm{n}(\bm{x})-\bm{n}(\bm{y})}{\lVert\bm{n}(\bm{x})-\bm{n}(\bm{y})\rVert},
  \label{eq:nsym_def}
\end{equation}
with a fallback to the local line-of-centers direction when $\lVert\bm{n}(\bm{x})-\bm{n}(\bm{y})\rVert$ is too small.
Finally, we orient $\bm{n}_{\mathrm{sym}}$ consistently so that it points from $\bm{x}$ toward the opposing interface (the partner cell),
which guarantees that the partner experiences the opposite direction when the same construction is applied at $\bm{y}$.
This choice (i) suppresses spurious tangential components and (ii) enhances antisymmetry of the interaction.

\textit{Repulsive flux and force density.}
We write the near-contact interaction as a repulsive ``flux'' $\bm{J}_{\mathrm{rep}}$ and the corresponding body-force density
\begin{equation}
  \bm{F}_{\mathrm{rep}}(\bm{x}) = \rho(\phi(\bm{x}))\,\bm{J}_{\mathrm{rep}}(\bm{x}).
\end{equation}
The closure adopted in this work is
\begin{equation}
  \bm{J}_{\mathrm{rep}}(\bm{x})
  =
  A_{\mathrm{rep}}\;
  \chi_{\mathrm{nc}}(\bm{x})\;
  q_{\mathrm{pair}}(\bm{x})\;
  w_h\!\big(h(q_{\mathrm{pair}})\big)\;
  f_{\mathrm{face}}(\bm{x})\;
  \bm{n}_{\mathrm{sym}}(\bm{x}).
  \label{eq:Jrep_final}
\end{equation}
Here $A_{\mathrm{rep}}$ is a tunable amplitude controlling the onset of non-coalescence, $h(q_{\mathrm{pair}})$ is an effective film-thickness indicator obtained in closed form from the phase-field profile, as derived in Sec.~\ref{sec:film_thickness_map}, and $w_h(h)$ is a bounded monotone weight that increases the repulsion as the effective interface separation decreases. A convenient and robust choice is
\begin{equation}
  w_h(h)=\frac{1}{1+(h/h_0)^p},
  \qquad h_0=\mathcal{O}(\varepsilon),\quad p\ge 2,
  \label{eq:wh_general}
\end{equation}
while our reference implementation uses $p=4$ and $h_0$ fixed in lattice units, i.e.\ $w_h(h)=1/(1+h^4)$.

Equation~\eqref{eq:Jrep_final} makes explicit the intended separation of roles:
$q_{\mathrm{pair}}$ localizes to the interface, $f_{\mathrm{face}}$ ensures that only opposing sheets interact, $w_h(h)$ concentrates
the interaction in thin films, and $\bm{n}_{\mathrm{sym}}$ enforces a 
predominantly normal and antisymmetric response.

\subsubsection{From phase-field overlap to an effective film-thickness indicator}
\label{sec:film_thickness_map}

A central aspect of the present approach is that a local measure of interface proximity is obtained in closed form from the phase-field indicator, without requiring ray tracing along the interface normal. The resulting quantity should be interpreted as an effective film-thickness indicator rather than as a direct geometric measurement of the actual separation between the interfaces. The construction relies on the equilibrium one-dimensional phase-field profile
\begin{equation}
  \phi(s)
  =
  \frac{1}{2}
  \left[
    1+\tanh\!\left(\frac{2s}{\varepsilon}\right)
  \right],
  \label{eq:tanh_profile}
\end{equation}
where $s$ denotes the signed distance from the center of the diffuse interface and $\varepsilon$ is the prescribed interface thickness.

\textit{Step 1: indicator profile along a planar interface.}
From $q(\phi)=\phi(1-\phi)$ and \eqref{eq:tanh_profile}, we obtain
\begin{equation}
  q(s)
  =
  \frac{1}{4}
  \left[
    1-\tanh^2\!\left(\frac{2s}{\varepsilon}\right)
  \right]
  =
  \frac{1}{4}\,
  \mathrm{sech}^2\!\left(\frac{2s}{\varepsilon}\right),
  \label{eq:q_of_s}
\end{equation}
which reaches its maximum value, $q=1/4$, at $s=0$ and decays exponentially as $|s|/\varepsilon$ increases.

\textit{Step 2: inversion $q\mapsto s$.}
From \eqref{eq:q_of_s},
\[
  \mathrm{sech}^2\!\left(\frac{2s}{\varepsilon}\right)=4q,
  \qquad
  \cosh\!\left(\frac{2s}{\varepsilon}\right)=\frac{1}{2\sqrt{q}},
\]
hence
\begin{equation}
  s(q)
  =
  \frac{\varepsilon}{2}\,\mathrm{arcosh}\!\left(\frac{1}{2\sqrt{q}}\right),
  \qquad 0<q\le\frac14.
  \label{eq:s_of_q}
\end{equation}

\textit{Step 3: symmetric thin-film thickness.}
In a symmetric near-contact film formed by two opposing interfaces, the (half-)gap is approximated by $s$, and the film thickness is
\begin{equation}
  h(q)\approx 2\,s(q)
  =
  \varepsilon\,\mathrm{arcosh}\!\left(\frac{1}{2\sqrt{q}}\right).
  \label{eq:h_of_q}
\end{equation}
In our model we evaluate \eqref{eq:h_of_q} using $q=q_{\mathrm{pair}}$ from \eqref{eq:q_pair}.
We emphasize that \eqref{eq:h_of_q} is intended to be used only within activated near-contact regions ($\chi_{\mathrm{nc}}=1$);
outside such regions, $q$ simply identifies an isolated diffuse interface and does not represent a physical film thickness.

\textit{Consistency and limitations.}
As $q\to 1/4^{-}$ (approaching an interface mid-surface), $h\to 0$ smoothly, while $h\to\infty$ as $q\to 0^{+}$ (bulk), in agreement
with the interpretation of $q$ as an overlap/thickness proxy in near-contact configurations.

\subsubsection{Divergence-form (stress) representations and planar-limit disjoining pressure}
\label{sec:stress_forms}

For the sake of interpretability, it is useful to express $\bm{F}_{\mathrm{rep}}$ as a divergence of a stress tensor,
$\bm{F}_{\mathrm{rep}}=\nabla\!\cdot\bm{\Sigma}_{\mathrm{rep}}$ (at least approximately). 

For the equilibrium profile \eqref{eq:tanh_profile}, one has
\begin{equation}
  \frac{d\phi}{ds}
  =
  \frac{1}{\varepsilon}\,\mathrm{sech}^2\!\left(\frac{2s}{\varepsilon}\right),
\end{equation}
and, using \eqref{eq:q_of_s},
\begin{equation}
  q(s) = \frac14\,\mathrm{sech}^2\!\left(\frac{2s}{\varepsilon}\right)
  =
  \frac{\varepsilon}{4}\,\left|\frac{d\phi}{ds}\right|.
  \label{eq:q_grad_identity_1d}
\end{equation}
For a locally planar interface $\nabla\phi=(d\phi/ds)\,\bm{n}$, and therefore
\begin{equation}
  q(\phi)\,\bm{n} = \frac{\varepsilon}{4}\,\nabla\phi.
  \label{eq:q_n_equals_gradphi}
\end{equation}
This identity is exact for the 1-D equilibrium profile and provides a direct route to a repulsive pressure interpretation.

\textit{Exact isotropic pressure form in the locally planar, equilibrium limit.}
Assume locally planar equilibrium conditions and suppose that the thickness weight can be expressed as a function of $\phi$ along the
profile, i.e.\ $w_h(h(q))\equiv W(\phi)$ locally.
Neglecting for the moment the facing modulation and activation (which take unit values in a perfectly symmetric film), the repulsive force
\eqref{eq:Jrep_final} reduces to
\begin{equation}
  \bm{F}_{\mathrm{rep}}
  =
  \rho\,A_{\mathrm{rep}}\,q(\phi)\,W(\phi)\,\bm{n}.
\end{equation}
Using \eqref{eq:q_n_equals_gradphi},
\begin{equation}
  \bm{F}_{\mathrm{rep}}
  =
  \rho\,A_{\mathrm{rep}}\,\frac{\varepsilon}{4}\,W(\phi)\,\nabla\phi
  =
  \nabla \Pi(\phi),
\end{equation}
where the scalar potential $\Pi(\phi)$ is
\begin{equation}
  \bm{F}_{\mathrm{rep}} = \nabla\!\cdot\big(\Pi(\phi)\,\bm{I}\big),
  \qquad
  \Pi(\phi)
  =
  A_{\mathrm{rep}}\,\frac{\varepsilon}{4}\int^{\phi}\rho(\xi)\,W(\xi)\,d\xi.
\label{eq:Pi_isotropic}
\end{equation}
Thus, in the locally planar equilibrium limit, the near-contact repulsion is exactly equivalent to the gradient of an isotropic, positive 
``disjoining pressure'' $\Pi(\phi)$.

\textit{General normal-stress form away from the planar limit.}
Away from the planar limit (curvature, non-equilibrium, spatial variability), a natural stress representation is a predominantly normal
interfacial stress
\begin{equation}
  \bm{F}_{\mathrm{rep}}
  =
  \nabla\!\cdot\!\Big(\Pi_{\mathrm{nc}}\,\bm{n}_{\mathrm{sym}}\otimes\bm{n}_{\mathrm{sym}}\Big),
\label{eq:anisotropic_stress}
\end{equation}
where $\Pi_{\mathrm{nc}}(\bm{x})$ is an effective near-contact normal stress amplitude.
Decomposing \eqref{eq:anisotropic_stress} along the normal coordinate $s$ associated with $\bm{n}_{\mathrm{sym}}$ and denoting by
$\kappa=\nabla\!\cdot\bm{n}_{\mathrm{sym}}$ the mean curvature of the normal field, one obtains (formally)
\begin{equation}
  \nabla\!\cdot(\Pi_{\mathrm{nc}}\bm{n}_{\mathrm{sym}}\otimes\bm{n}_{\mathrm{sym}})
  =
  \big(\partial_s\Pi_{\mathrm{nc}}+\kappa\,\Pi_{\mathrm{nc}}\big)\,\bm{n}_{\mathrm{sym}}
  + \Pi_{\mathrm{nc}}(\bm{n}_{\mathrm{sym}}\!\cdot\!\nabla)\bm{n}_{\mathrm{sym},\perp},
  \label{eq:div_nn_general}
\end{equation}
where the second term is tangential and vanishes for planar configurations (and is $\mathcal{O}(\kappa)$ for gentle curvature).
Matching the leading normal component of \eqref{eq:div_nn_general} to the model forcing direction in \eqref{eq:Jrep_final} yields a
normal-balance relation for $\Pi_{\mathrm{nc}}$:
\begin{equation}
  \partial_s\Pi_{\mathrm{nc}}+\kappa\,\Pi_{\mathrm{nc}}
  \;\approx\;
  \rho\,A_{\mathrm{rep}}\,
  \chi_{\mathrm{nc}}\,
  q_{\mathrm{pair}}\,
  w_h(h(q_{\mathrm{pair}}))\,
  f_{\mathrm{face}}.
  \label{eq:Pi_ode_general}
\end{equation}
In the locally planar case ($\kappa\simeq 0$), \eqref{eq:Pi_ode_general} reduces to $\partial_s\Pi_{\mathrm{nc}}\approx(\cdots)$ and the
isotropic disjoining-pressure picture \eqref{eq:Pi_isotropic} is recovered when the dependence can be expressed through $\phi$ along an
equilibrium profile.

The construction \eqref{eq:Jrep_final} is designed to act predominantly along $\bm{n}_{\mathrm{sym}}$.
If a strictly tangential-free representation is desired at the discrete level, one may also consider adding an isotropic counter-stress
on the tangent plane. In practice, we find that the symmetric-normal choice already suppresses spurious tangential contributions and
provides robust near-contact stabilization.

\subsubsection{Summary of the closure and model parameters}
\label{sec:summary_params}

Collecting the definitions, the near-contact interaction entering \eqref{eq:mom} is
\begin{equation}
  \bm{F}_{\mathrm{rep}}(\bm{x})
  =
  \rho(\phi(\bm{x}))\,
  A_{\mathrm{rep}}\,
  \chi_{\mathrm{nc}}(\bm{x})\,
  q_{\mathrm{pair}}(\bm{x})\,
  w_h\!\big(h(q_{\mathrm{pair}})\big)\,
  f_{\mathrm{face}}(\bm{x})\,
  \bm{n}_{\mathrm{sym}}(\bm{x}),
\end{equation}
with $q_{\mathrm{pair}}$ from \eqref{eq:q_pair}, $f_{\mathrm{face}}$ from \eqref{eq:facing_factor},
$\bm{n}_{\mathrm{sym}}$ from \eqref{eq:nsym_def}, and $h(q)$ from \eqref{eq:h_of_q}.

\begin{figure}
    \centering
    \includegraphics[width=0.95\linewidth]{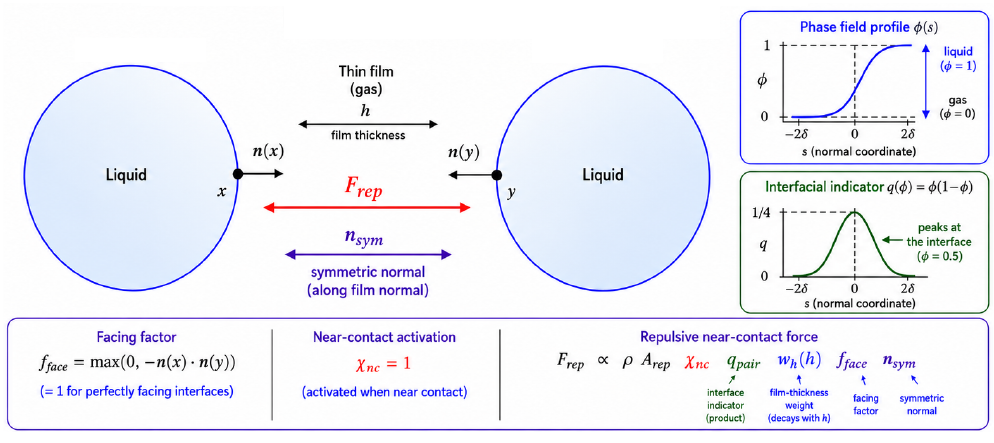}
    \caption{
Schematic of the adaptive near-contact interaction model. The phase field distinguishes the liquid phase, $\phi=1$, from the surrounding bulk phase, $\phi=0$. When two interfaces face each other across an unresolved gas film of thickness $h$, a local repulsive force $\mathbf{F}_{\mathrm{rep}}$ is activated along the symmetric film-normal direction $\mathbf{n}_{\mathrm{sym}}$, with strength controlled by the interfacial indicator $q(\phi)$, the facing factor $f_{\mathrm{face}}$, and the film-thickness weight $w_h(h)$.
}
    \label{fig:nci_sketch}
\end{figure}

The free parameters are:
(i) $A_{\mathrm{rep}}$, controlling the onset of non-coalescence and the overall strength of the near-contact barrier;
(ii) the thickness-weight parameters $(h_0,p)$ in \eqref{eq:wh_general}, setting how sharply repulsion concentrates as films thin;
(iii) the activation/pairing thresholds implicitly contained in $\chi_{\mathrm{nc}}$ (interfacial gate, facing threshold, neighborhood
radius). Their discrete construction is reported in Sec.~\ref{sec:nci_implementation}; importantly, the method avoids ray tracing along interface normals and instead combines bounded local detection with the local proximity mapping defined in \eqref{eq:h_of_q}. A schematic representation of the near-contact repulsive model is provided in Fig.~\ref{fig:nci_sketch}.

\textcolor{black}{In the present model, the amplitude $A_{\mathrm{rep}}$ is chosen to be commensurate with the capillary-force scale, typically $A_{\mathrm{rep}}=\mathcal{O}(\sigma)$. Once the non-coalescing regime is reached, the adaptive response discussed in Sec.~\ref{sec:self_tuning} considerably reduces the sensitivity to its precise value. The parameters $h_0$ and $p$ determine the range and sharpness of the thickness-dependent weight $w_h(h)=[1+(h/h_0)^p]^{-1}$. Here, $h_0=1$ lattice unit and $p=4$, confining appreciable forcing to effective gaps of order one lattice spacing. In general, $h_0$ should reflect the smallest unresolved film thickness to be stabilized, whereas values $p\geq2$ yield a smooth and localized response, with larger $p$ producing sharper localization.}
%===========================================================

%===========================================================
\subsection{Numerical implementation of the near-contact interaction}
\label{sec:nci_implementation}

This section describes the lattice implementation of the near-contact interaction (NCI). The procedure is based on a bounded local stencil, is well suited to GPU implementation, and is designed to activate only for facing-interface configurations associated with thin intervening films, without requiring ray tracing or marching along interface normals. At each time step, the implementation consists of two stages: (1) \emph{near-contact detection and pairing}, which constructs an activation mask and a reciprocal local partner map, and (2) \emph{flux/force construction}, which evaluates the closed-form phase-field-based proximity indicator and returns a repulsive force aligned with the symmetrized interface normal. The pairing and force construction are designed to guarantee discrete pairwise antisymmetry, so that each interacting pair receives equal and opposite force contributions.
%-----------------------------------------------------------
\subsubsection{Discrete geometric fields and interfacial gate}
\label{sec:nci_geom}

We assume that the phase field $\phi$ is available at cell centers. 
We first compute the interface normal field
\begin{equation}
  \bm{n}(\bm{x})=\frac{\nabla\phi(\bm{x})}{\lVert\nabla\phi(\bm{x})\rVert+\epsilon_n},
  \label{eq:n_discrete}
\end{equation}
where $\epsilon_n$ is a small regularization parameter. 
In practice, $\nabla\phi$ is evaluated with an
isotropic stencil consistent with the lattice (e.g.\ D3Q27 isotropic gradient) 
to minimize directional bias.

We also compute the interfacial indicator
\begin{equation}
  q(\bm{x}) = \phi(\bm{x})\bigl(1-\phi(\bm{x})\bigr),
  \label{eq:q_discrete}
\end{equation}
with $\phi$ clamped to $[0,1]$ when needed (as in the code) to avoid small overshoots affecting the gate.
A cell is considered \emph{interfacial} if
\begin{equation}
  q(\bm{x}) \ge q_{\mathrm{th}},
  \label{eq:q_gate}
\end{equation}
where $q_{\mathrm{th}}$ is a small threshold (typically $q_{\mathrm{th}}\ll 1/4$).
This gate restricts the search to cells located within the diffuse interfacial band.

%-----------------------------------------------------------
\subsubsection{Thin-film detection and partner selection}
\label{sec:nci_detection}

\textit{Overview.}
The purpose of detection is to activate the NCI on demand, i.e. 
only in regions where two interfaces are in near contact \emph{and} face each other.
At each cell $\bm{x}$ satisfying \eqref{eq:q_gate}, we search in a bounded cubic neighborhood (window radius $w$):
\begin{equation}
  \mathcal{W}(\bm{x})=\{\bm{x}+\bm{r}\;:\; \bm{r}=(d_i,d_j,d_k),\ d_\alpha\in[-w,w],\ \bm{r}\neq \bm{0}\},
  \label{eq:window}
\end{equation}
and attempt to identify a \emph{partner} cell $\bm{y}\in \mathcal{W}(\bm{x})$ located on the opposing interface across the film.

We store:
(i) a binary activation mask $\chi_{\mathrm{nc}}(\bm{x})\in\{0,1\}$ ,
and (ii) an integer partner map $\bm{x}_p(\bm{x})$.

\textit{Candidate acceptance criteria.}
A neighbor cell $\bm{y}\in\mathcal{W}(\bm{x})$ qualifies as a valid 
partner candidate if it satisfies three conditions:

\begin{enumerate}
\item \textbf{Interfacial neighbor gate:} $q(\bm{y})\ge q_{\mathrm{th}}$.

\item \textbf{Iso-interface similarity:}
\begin{equation}
  |q(\bm{y})-q(\bm{x})|
  \le \eta\ \max\!\bigl(q(\bm{x}),\epsilon_q\bigr),
  \label{eq:q_similarity}
\end{equation}
with $\eta=0.1$ in our implementation and $\epsilon_q$ a small floor.
This enforces that the two points lie at comparable locations within the diffuse interface (e.g.\ both close to $\phi\simeq 1/2$),
which improves symmetry and reduces false positives where an interfacial cell pairs with a near-bulk cell.

\item \textbf{Facing condition (opposing normals):}
\begin{equation}
  \bm{n}(\bm{x})\cdot\bm{n}(\bm{y}) \le \cos\theta_{\mathrm{opp}},
  \label{eq:facing_threshold}
\end{equation}
where $\theta_{\mathrm{opp}}$ is an ``opposition'' angle (e.g.\ close to $\pi$). 
This prevents activation on the same interface sheet.
\end{enumerate}

In addition, for each accepted candidate, we define a \emph{facing score} used both for tie-breaking and later for the force magnitude.
Consistent with the forcing stage, we use
\begin{equation}
  f_{\mathrm{face}}(\bm{x},\bm{y})=\max\!\Bigl(0,\ -\bm{n}(\bm{x})\cdot\bm{n}(\bm{y})\Bigr)\in[0,1].
  \label{eq:facing_score}
\end{equation}
(Any equivalent monotone mapping of the dot product can be used; our code, \textit{accLB}, uses an affine mapping for selection and
\eqref{eq:facing_score} for the final forcing, which are consistent up to monotone rescaling.)

\textit{Nearest-partner selection rule.}
Among all accepted candidates, we choose the partner $\bm{y}_{\mathrm{best}}$ as the \emph{nearest} one in Euclidean distance,
\begin{equation}
  \bm{y}_{\mathrm{best}} = \arg\min_{\bm{y}\in\mathcal{C}(\bm{x})} \ |\bm{y}-\bm{x}|^2,
\end{equation}
where $\mathcal{C}(\bm{x})$ is the set of accepted candidates.
If multiple candidates share the same minimal distance (ties), we choose the one with the largest facing score
$f_{\mathrm{face}}(\bm{x},\bm{y})$.
The resulting activation mask and partner map are then defined as
\begin{equation}
  \chi_{\mathrm{nc}}(\bm{x})=
  \begin{cases}
    1, & \text{if at least one candidate exists},\\
    0, & \text{otherwise},
  \end{cases}
  \qquad
  \bm{x}_p(\bm{x})=
  \begin{cases}
    \bm{y}_{\mathrm{best}}, & \chi_{\mathrm{nc}}(\bm{x})=1,\\
    \bm{0}, & \chi_{\mathrm{nc}}(\bm{x})=0.
  \end{cases}
  \label{eq:mask_partner_def}
\end{equation}

\textit{Remarks on locality and robustness.}
The method relies on a bounded local search window and is therefore well suited to GPU implementation. For a fixed search radius $w$, the detection cost scales as $\mathcal{O}\!\left(N(2w+1)^3\right)$. The facing criterion in \eqref{eq:facing_threshold} is used to distinguish a genuine thin-film configuration, formed by two opposing diffuse interfaces, from a single isolated interface, where nearby interfacial points instead have similarly oriented normals. The similarity gate in \eqref{eq:q_similarity} further favors the pairing of points located at comparable depths within the two diffuse interfaces, thereby improving the consistency of the phase-field-based proximity indicator. Finally, the reciprocal partner map and the symmetrized interface normal employed in the force construction guarantee discrete pairwise antisymmetry, so that each interacting pair receives equal and opposite force contributions.

\textcolor{black}{\textit{Parameter selection.} In all the simulations presented here, the search-window radius is set to $w=3$ lattice nodes, the interfacial threshold to $q_{\mathrm{th}}=0.125$, the iso-interface tolerance to $\eta=0.1$, and the normal-opposition threshold to $\cos\theta_{\mathrm{opp}}=-0.8$. These parameters govern the identification of admissible near-contact pairs rather than the magnitude of the repulsive force. The choice $w=3$ defines a search window extending over $2w+1=7$ lattice nodes in each direction, consistently with the diffuse-interface thickness of approximately six lattice nodes adopted in the present simulations. The window is therefore large enough to detect an opposing interface across an unresolved film while retaining a compact local stencil. The value $q_{\mathrm{th}}=0.125=q_{\max}/2$, with $q_{\max}=1/4$, confines the search to the central portion of the diffuse interface and excludes near-bulk cells. The iso-interface tolerance promotes pairing between points located at comparable positions within the two diffuse-interface profiles, while the condition $\bm{n}(\bm{x})\cdot\bm{n}(\bm{y})\leq-0.8$ selects clearly opposing interface orientations and prevents activation between neighboring points belonging to the same interface sheet. The influence of marginal candidate configurations is further regularized by the continuous factors $q_{\mathrm{pair}}$, $f_{\mathrm{face}}$, and $w_h(h)$, which smoothly modulate the resulting force. These detection parameters are distinct from the force-amplitude and thickness-weight parameters discussed in Sec.~\ref{sec:summary_params}.}

\textcolor{black}{\textit{Computational cost and scalability.} The main additional cost of the near-contact treatment is associated with the bounded neighborhood search used for interface pairing. For a search-window radius $w$, at most $(2w+1)^3-1$ neighboring positions are inspected for each searched cell. In the present simulations, $w=3$, corresponding to a maximum neighborhood of $7^3-1=342$ candidate positions. The search is performed only for cells satisfying the interfacial condition $q\geq q_{\mathrm{th}}=0.125$, rather than over the full computational domain. The effective cost therefore scales primarily with the instantaneous exposed interfacial area: the overhead is lowest in dilute systems and increases with the number and total area of the interfaces. The most demanding configuration considered here is the rising-bubble swarm, which contains approximately one hundred bubbles and undergoes repeated many-body near-contact events. In this case, the NCI treatment introduces an overhead of approximately $10\%$ relative to a simulation performed with the same multiphase solver but without NCI. Once a partner has been identified, evaluating $q_{\mathrm{pair}}$, $f_{\mathrm{face}}$, $h(q_{\mathrm{pair}})$, $w_h(h)$, and $\bm{n}_{\mathrm{sym}}$ requires only local algebraic operations and adds a comparatively small cost. For fixed $w$, the complete procedure retains $\mathcal{O}(N)$ complexity with the number of lattice nodes and requires neither ray tracing nor global geometric reconstruction, iterative film solvers, or additional global communication. It therefore preserves the locality and parallel scalability characteristic of lattice Boltzmann implementations, a computational paradigm that has enabled simulations at extreme spatial scales on massively parallel architectures~\cite{falcucci2021extreme}.}

%-----------------------------------------------------------
\subsubsection{Repulsive forcing construction from the partner map}
\label{sec:nci_force_construction}

Once $\chi_{\mathrm{nc}}$ and $\bm{x}_p$ are available, we compute a purely normal repulsive forcing at each activated cell.

\textit{Paired indicator and clamping.}
Let $\bm{y}=\bm{x}_p(\bm{x})$ for an activated $\bm{x}$. We define
\begin{equation}
  q_1=q(\bm{x}),\qquad q_2=q(\bm{y}),\qquad q_{\mathrm{pair}}=\tfrac12(q_1+q_2).
  \label{eq:qpair_impl}
\end{equation}
To avoid numerical issues in the film-thickness mapping (which involves $\sqrt{q}$ and $\mathrm{arcosh}$), we clamp
\begin{equation}
  q_{\mathrm{cl}} = \min\!\bigl(\max(q_{\mathrm{pair}},\epsilon_q),\ 1/4-\epsilon_q\bigr),
  \label{eq:q_clamp}
\end{equation}
with $\epsilon_q$ a small number.

\textit{Facing factor and symmetric normal direction.}
We compute the facing factor (same as in the theory section)
\begin{equation}
  f_{\mathrm{face}}=\max\!\Bigl(0,\ -\bm{n}(\bm{x})\cdot\bm{n}(\bm{y})\Bigr).
  \label{eq:face_impl}
\end{equation}
If $f_{\mathrm{face}}$ is zero, the interaction is suppressed.

We then build the symmetric normal direction
\begin{equation}
  \tilde{\bm{n}}_{\mathrm{sym}} = \bm{n}(\bm{x})-\bm{n}(\bm{y}),
  \qquad
  \bm{n}_{\mathrm{sym}}=
  \frac{\tilde{\bm{n}}_{\mathrm{sym}}}{\lVert\tilde{\bm{n}}_{\mathrm{sym}}\rVert+\epsilon_n},
  \label{eq:nsym_impl}
\end{equation}
with fallback to the line-of-centers unit vector
\[
  \hat{\bm{r}}=\frac{\bm{y}-\bm{x}}{\lVert\bm{y}-\bm{x}\rVert}
\]
when $\lVert\tilde{\bm{n}}_{\mathrm{sym}}\rVert$ is too small.
Finally, we orient $\bm{n}_{\mathrm{sym}}$ so that $\hat{\bm{r}}\cdot\bm{n}_{\mathrm{sym}}\ge 0$.
This convention ensures that the partner cell, when processed, produces the opposite direction, enhancing action--reaction behavior.

\textit{Closed-form film-thickness indicator and thickness weight.}
A key feature of the present approach is that an effective local film-thickness indicator can be obtained in closed form from $q_{\mathrm{cl}}$. By inverting the equilibrium-profile relation derived in Sec.~\ref{sec:film_thickness_map}, we compute
\begin{equation}
  h
  =
  \varepsilon\,\mathrm{arcosh}\!\left(
  \frac{1}{2\sqrt{q_{\mathrm{cl}}}}
  \right),
  \label{eq:h_impl}
\end{equation}
where $\varepsilon$ is the diffuse-interface thickness scale. The quantity $h$ is therefore interpreted as a phase-field-based indicator of local interface proximity, calibrated on the equilibrium one-dimensional profile, rather than as a direct geometric measurement of the actual film thickness.

The corresponding bounded thickness weight is then evaluated as $w_h(h)$. In the reference implementation, we use
\begin{equation}
  w_h(h)
  =
  \frac{1}{1+h^4},
  \label{eq:wh_impl}
\end{equation}
which corresponds to the general form
$w_h(h)=1/[1+(h/h_0)^p]$ with $p=4$ and $h_0=1$ in lattice units, consistently with the parameter choice discussed in Sec.~\ref{sec:summary_params}.

\textit{Final repulsive flux and force.}
With these ingredients, the discrete repulsive flux is
\begin{equation}
  \bm{J}_{\mathrm{rep}}(\bm{x})
  =
  A_{\mathrm{rep}}\;
  \chi_{\mathrm{nc}}(\bm{x})\;
  q_{\mathrm{cl}}(\bm{x})\;
  w_h(h)\;
  f_{\mathrm{face}}(\bm{x})\;
  \bm{n}_{\mathrm{sym}}(\bm{x}),
  \label{eq:Jrep_impl}
\end{equation}
and the force density entering the momentum equation is
\begin{equation}
  \bm{F}_{\mathrm{rep}}(\bm{x})=\rho(\phi(\bm{x}))\,\bm{J}_{\mathrm{rep}}(\bm{x}).
  \label{eq:Frep_impl}
\end{equation}

\section{Results}

\subsection{Head-on and off-axis impacts between non-coalescing droplets: comparison with experiments} \label{sec:exp_comparison}

\begin{figure}
    \centering
    \includegraphics[width=0.75\linewidth]{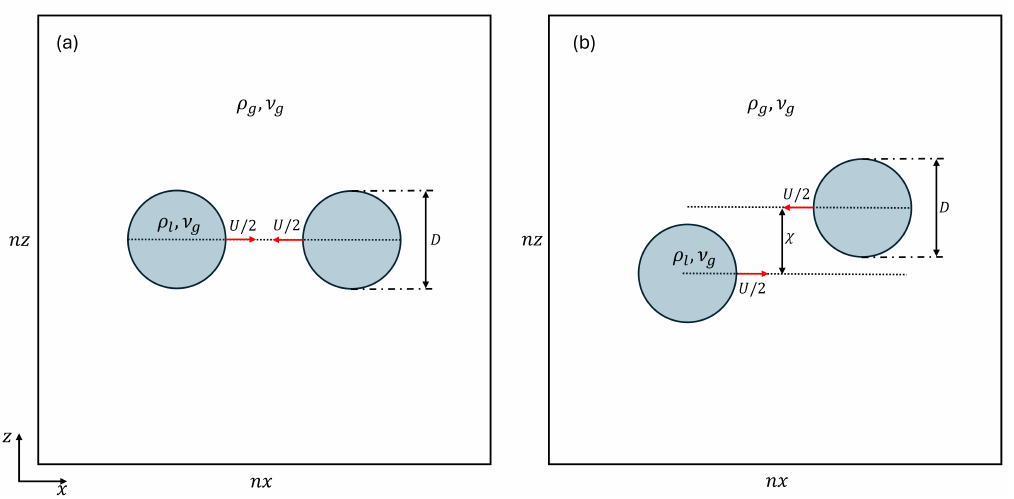}
    \caption{Graphical sketch of the head-on (a) and off-axis (b) collisions between equally sized droplets.}
    \label{fig:sketchdrop}
\end{figure}

\begin{figure}
    \centering
    \includegraphics[width=0.9\linewidth]{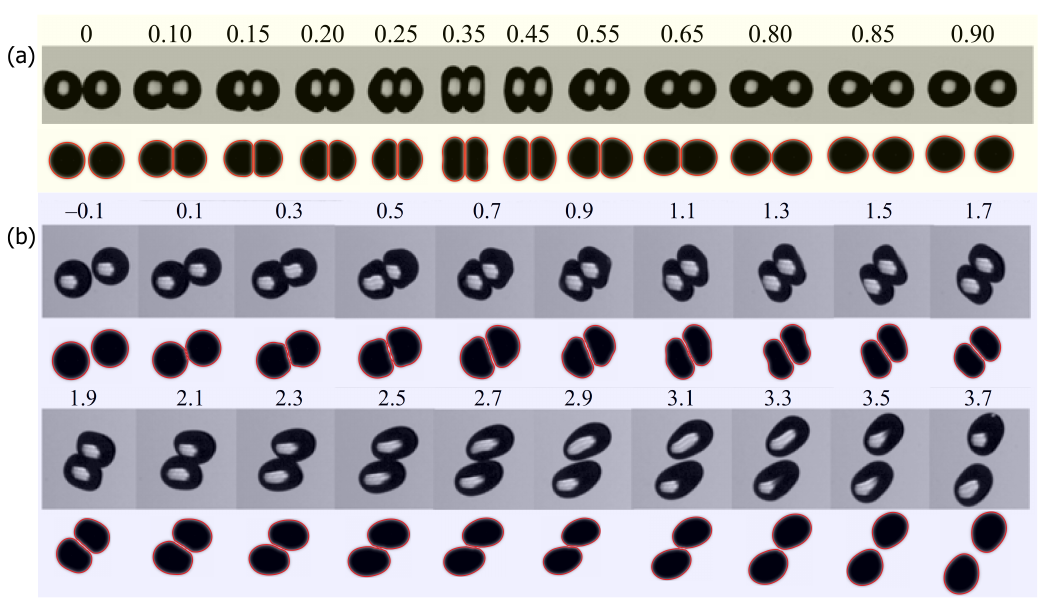}
    \caption{Comparison between experimental results of Huang \& Pan~\cite{HuangPan2021JFM} and the corresponding numerical simulations for two bouncing regimes. (a) Head-on collision of dodecane droplets with $D=300~\mu\mathrm{m}$ at $We = 4.60$. (b) Off-center collision of water droplets with $D=1000~\mu\mathrm{m}$, $ B = 0.32$ and $We = 4.8$. In each case, experiments (top) and simulations (bottom) are shown for direct comparison. The time shown above each frame is given in milliseconds.}
    \label{fig:exp_vs_num}
\end{figure}

To validate the proposed near-contact interaction (NCI) in a regime where the collision outcome is controlled by the unresolved thin gas film, we reproduce two reference experiments from Huang \& Pan~\cite{HuangPan2021JFM} featuring clean droplet bouncing at comparable Weber numbers but different droplet sizes and impact parameters. Specifically, we consider (i) a head-on collision of dodecane droplets with diameter $D=300~\mu\mathrm{m}$ at $\mathrm{We}=4.60$ and $B=0$, and (ii) an off-axis collision of water droplets with diameter $D=1000~\mu\mathrm{m}$ at $\mathrm{We}=4.8$ and $B=0.32$~\cite{HuangPan2021JFM} (see Fig.~\ref{fig:sketchdrop} for a graphical sketch of the experimental setup). Here, the Weber number is defined as $\mathrm{We}=\rho_{\ell} U^{2} D/\sigma$, and the impact parameter as $B=\chi/D$, with $\chi$ the offset between droplet centres measured in the direction normal to the relative velocity (collision plane)~\cite{HuangPan2021JFM}. These two cases (reported in Fig.~\ref{fig:exp_vs_num}) are particularly stringent benchmarks for diffuse-interface models because the physically relevant film thickness at rebound remains far below the computational interfacial thickness; without an additional short-range stabilization mechanism, premature interface overlap typically leads to spurious coalescence.
All simulations are performed in a fully periodic cubic domain of size $512^{3}$ lattice units (lu). Each droplet has diameter $D_{\mathrm{lu}}=80$, which sets the spatial mapping $\Delta x = D/D_{\mathrm{lu}}$; this yields $\Delta x=3.75~\mu\mathrm{m}$ for the $D=300~\mu\mathrm{m}$ head-on case and $\Delta x=12.5~\mu\mathrm{m}$ for the $D=1000~\mu\mathrm{m}$ off-axis case. The density ratio is fixed to $\rho_{\ell}/\rho_{g}=1000$, and the surface tension is set to $\sigma=0.02$ (lu). Kinematic viscosities are prescribed in lattice units and selected so that the simulations match the experimental nondimensional conditions (in particular $\mathrm{We}$, and the associated inertio-capillary time scale): for the head-on case we use $\nu_{\ell}=0.005$ and $\nu_{g}=0.05$ (lu), whereas for the off-axis case $\nu_{\ell}=0.0012$ and $\nu_{g}=0.012$ (lu). In all comparisons, the reported time is expressed in milliseconds, consistently with the experimental sequences~\cite{HuangPan2021JFM}.

Fig.~\ref{fig:exp_vs_num} provides a direct visual comparison between the experimental silhouettes (top rows) and the numerical interfaces (bottom rows). In the head-on collision at $D=300~\mu\mathrm{m}$ and $\mathrm{We}=4.60$, the simulation reproduces the symmetric approach, the progressive flattening at impact, and the subsequent rebound without coalescence, with a close match of the deformation levels and of the rebound timing (the droplets detach at $t\simeq 0.85\text{--}0.90~\mathrm{ms}$ in both experiment and simulation). In the off-axis case ($D=1000~\mu\mathrm{m}$, $\mathrm{We}=4.8$, $B=0.32$), the model captures the oblique interaction sequence: the droplets undergo asymmetric deformation, exchange tangential momentum during near-contact, and then re-separate while preserving distinct interfaces, reproducing the experimentally observed post-impact shapes and the qualitative trajectory/rotation pattern over the full temporal window up to $t=3.7~\mathrm{ms}$. Minor discrepancies at late times are expected because the experimental images are 2D projections of a 3D event and because small differences in the initial alignment/relative phase can affect the long-time drift after rebound; nonetheless, the agreement across both a head-on and an off-center configuration indicates that the NCI provides the missing short-range stabilization required to recover the correct non-coalescing outcome within a diffuse-interface framework.

\subsection{Thin-film hydrodynamics during droplet impact}

\begin{figure}
    \centering
    \includegraphics[width=0.75\linewidth]{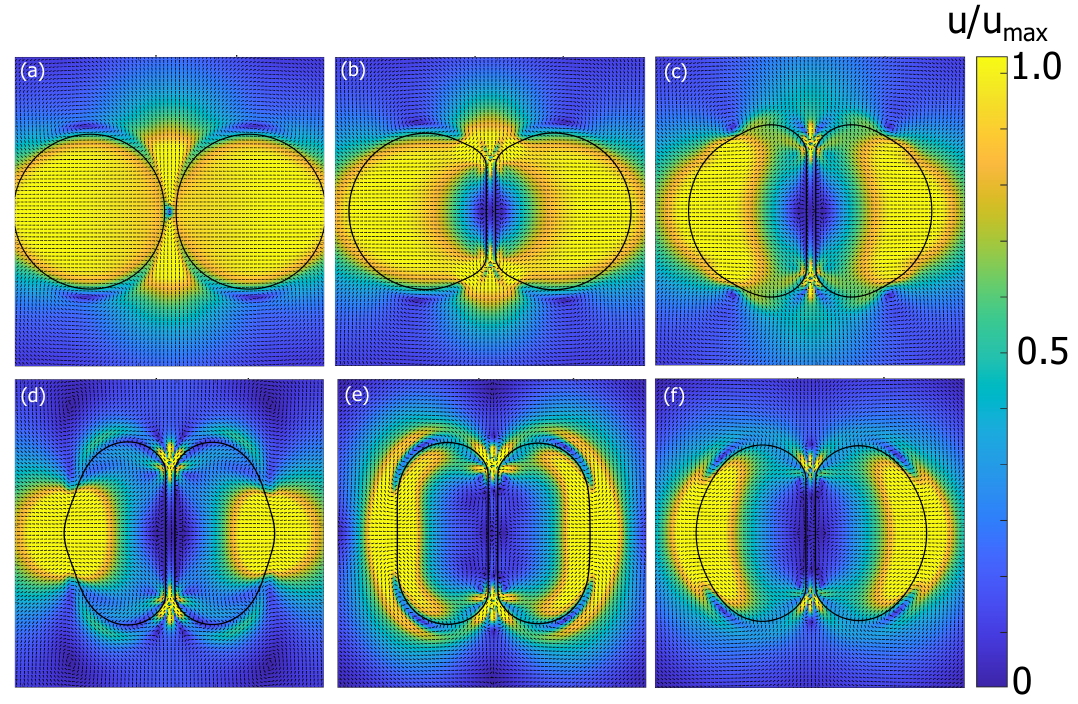}
    \caption{Velocity field in the inter-droplet region during head-on impact. The color map denotes the normalized velocity magnitude ($u/u_{\max}$), and arrows indicate the flow direction. 
    }
    \label{fig:vel_impact_headon}
\end{figure}

\textit{Flow-field signatures of thin-film sustenance and rebound (head-on case).}
To further substantiate that the proposed NCI reproduces not only the interface kinematics but also the
\emph{mechanism} responsible for non-coalescence, Fig.~\ref{fig:vel_impact_headon} reports a sequence of
velocity fields during the head-on impact (colormap: normalized velocity magnitude $u/u_{\max}$; arrows:
unit vectors, i.e. direction only). At early times (panel (a)), the flow in the inter-droplet film is intense
and predominantly directed \emph{outwards} along the collision plane, consistent with a lubrication-type
squeezing/draining flow driven by the rapid reduction of the gap. At the same stage, the velocity within each
droplet remains almost undisturbed: the core motion is essentially uniform and aligned with the impact direction
($x$-axis), indicating that the internal flow has not yet strongly reorganized in response to the impending
near-contact interaction. As the droplets reach the closest approach (panels (b)--(c)), the velocity within the thin
film \emph{drops sharply} (blue region in the colormap) and the in-plane flow direction in the film \emph{inverts},
highlighting the onset of a sustained interfacial separation. In parallel, the internal droplet flow departs from
the initial plug-like motion: the rear side of each droplet still moves towards the impact region, whereas near the
front (close to the interaction zone) the flow becomes predominantly tangential to the interface. This
reorientation is a clear signature of the pressure build-up in the film and of the associated redistribution of
momentum within the droplets, which is a necessary condition for rebound without coalescence. A distinctive
feature is observed at the \emph{rim} (tip) of the thin film (panels (c)--(d)): the velocity vectors display a
localized recirculation pattern, i.e. the flow is diverted and turns back rather than feeding through the film.

This is consistent with the fact that, during the interaction, the film behaves as a highly resistive lubrication
layer: the outward draining is progressively impeded, the fluid is redirected around the film edge, and the
resulting recirculation contributes to maintaining a finite separation up to the rebound stage. In the later
stages (panels (e)--(f)), the velocity within the film remains comparatively small, whereas the flow inside the droplets
\emph{reverses} and becomes directed outward with respect to the collision direction, consistent with the global
bouncing motion and the relaxation of the deformed interfaces after the closest approach. 

\begin{figure}
    \centering
    \includegraphics[width=0.75\linewidth]{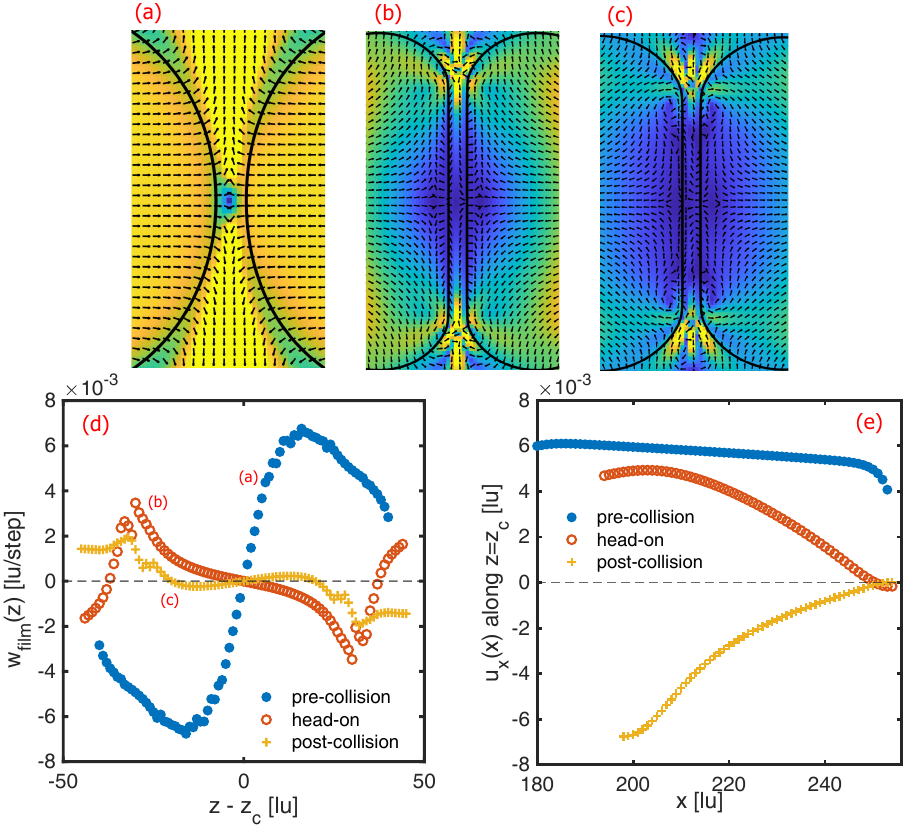}
    \caption{Velocity field and associated profiles at three stages of the collision. Panels (a)--(c) show the flow field at the pre-collision stage, at head-on contact, and in the post-collision stage. Panel (d) reports the film-averaged normal velocity $\bar{w}_{\mathrm{film}}$, and panel (e) shows the centerline streamwise velocity $u_x(x)$ along $z = z_0$. In panels (d) and (e), the blue, orange, and yellow curves correspond to the stages shown in panels (a)--(c), respectively.}
    \label{fig:wavg_film}
\end{figure}

% --- Local quantitative pointers (thin-film and centreline profiles) ---
To keep the above discussion quantitative, we report two local diagnostics extracted from the numerical fields. First, Fig.~\ref{fig:wavg_film} shows the film-averaged normal velocity $\bar{w}_{\mathrm{film}}(z)$, obtained by averaging $w$ across the instantaneous interfacial gap (identified between the two facing $\rho=\rho_{\mathrm{iso}}$ iso-contours) at each $z$. In the pre-impact stage, $\bar{w}_{\mathrm{film}}$ exhibits opposite signs on the two sides of the film centerline $z_0$: it is negative for $z>z_0$ and positive for $z<z_0$, consistently with a strong drainage pattern where the film flow is directed away from the interaction region with opposite orientation in the two semi-domains. As the droplets approach their minimum separation, the magnitude of $\bar{w}_{\mathrm{film}}$ collapses towards zero over most of the interaction region, indicating that the through-film flux is strongly reduced when the lubrication resistance becomes dominant. At later times, $\bar{w}_{\mathrm{film}}$ remains small and displays a reversal of the sign distribution compared with the early stage, providing a compact quantitative signature of the flow inversion in the thin film discussed above.
Second, Fig.~\ref{fig:wavg_film} reports the centerline profiles of the streamwise velocity $u_x(x)$ extracted along a fixed line $z=z_0$ inside the impacting droplet. In the pre-impact stage (panel (a)), $u_x$ is nearly constant throughout most of the droplet extent along $x$, indicating an almost plug-like internal motion aligned with the collision direction. During the interaction (panel (b)), the profile departs from this uniform state and progressively decreases towards the front region facing the thin film, approaching vanishing values in the vicinity of the interface. In the post-impact stage (panel (c)), this front-region attenuation persists and extends over a wider portion of the droplet, consistent with the emergence of a stagnation-like zone induced by the strong hydrodynamic resistance of the film and with a redistribution of momentum away from the incoming streamwise component. Together, these two diagnostics support the mechanistic picture inferred from the vector fields: the film drainage initially dominates, then the film flux is suppressed and reverses while the incoming axial motion is locally arrested near the impact region, enabling thin-film sustenance and subsequent rebound without coalescence.

\subsection{Self-tuning of the repulsive near-contact force}
\label{sec:self_tuning}
In this section, we report on an apparently counterintuitive feature observed when varying the near-contact repulsive amplitude $A_{\mathrm{rep}}$:
once $A_{\mathrm{rep}}$ exceeds the threshold required to prevent coalescence, further increases of $A_{\mathrm{rep}}$ produce
no discernible change in the macroscopic impact dynamics (e.g.\ rebound/approach kinematics, deformation patterns, and the overall
evolution of the colliding droplets).

\begin{figure}
    \centering
    \includegraphics[width=0.75\linewidth]{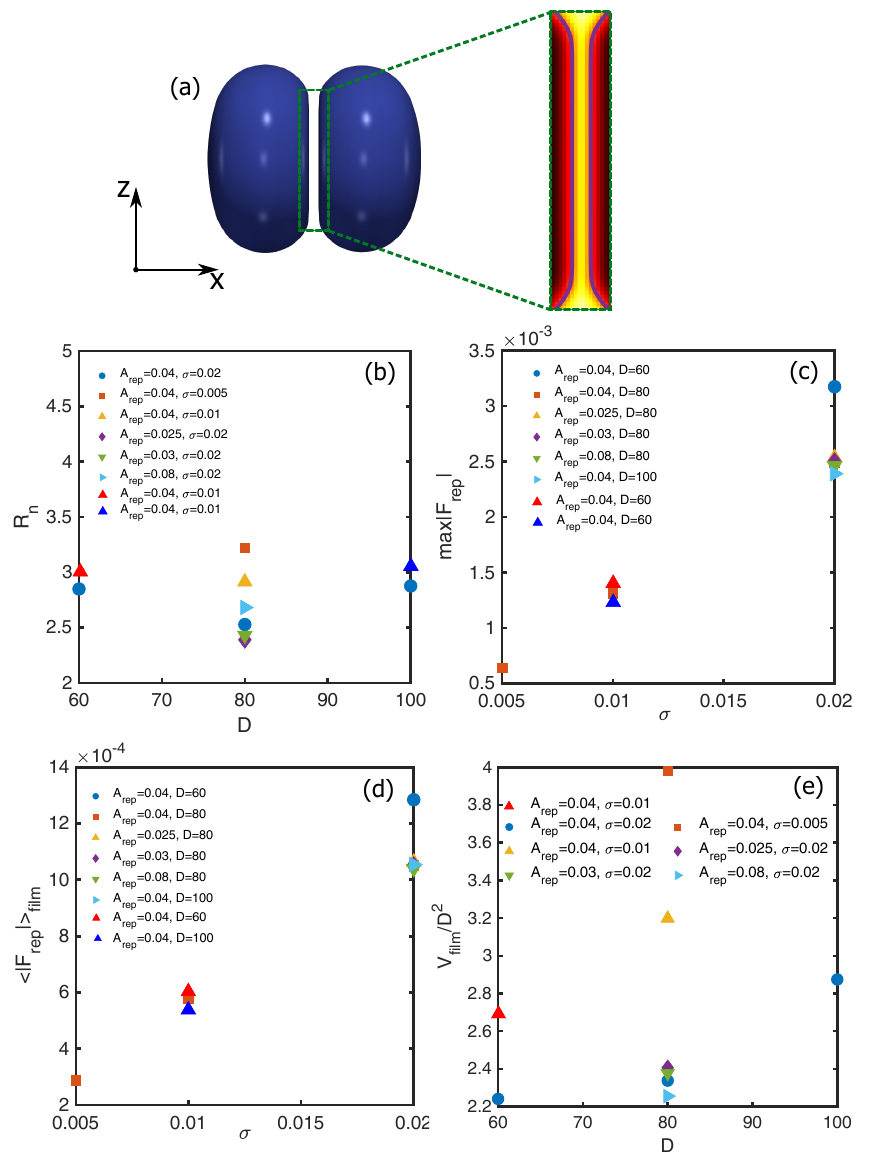}
    \caption{Quantification of the near-contact interaction and the capillary contribution within the inter-droplet film at peak activation for different values of $A_{\mathrm{rep}}$, $\sigma$, and $D$. (a) Schematic of the inter-droplet thin-film region where the force balance is evaluated. (b) Ratio between the repulsive and capillary forces within the thin film, $R_n$, as a function of droplet diameter. (c) Maximum repulsive force magnitude $\max|\mathbf{F}_{\mathrm{rep}}|$ as a function of $\sigma$. (d) Film-averaged repulsive force magnitude $\langle|\mathbf{F}_{\mathrm{rep}}|\rangle_{\mathrm{film}}$ as a function of $\sigma$. (e) Activated film volume normalized by the square of the droplet diameter $V_{\mathrm{film}}/D^2$ as a function of $D$.}
    \label{fig:repulsion_vs_capillary_plots}
\end{figure}
This observation suggests that the near-contact interaction does not 
act as a simple linear penalty factor (for which one would expect a
monotonic increase of repulsive stresses and a strong sensitivity of the dynamics to $A_{\mathrm{rep}}$), but rather as an
{\it adaptive barrier} that delivers the level of normal 
resistance required by the local thin-film force balance.
To test this hypothesis and rationalize the observed macroscopic robustness, we quantify the near-contact and capillary actions inside
the film by constructing a set of key observables computed at the instant of maximum activation (i.e.\ the time at which the
near-contact region is maximally activated). Specifically, we (i) compare the repulsive and capillary force components projected along
the local interface normal, (ii) monitor the absolute intensity of the repulsive forcing (peak and mean), and (iii) characterize how
the extent of the interaction region varies with droplet diameter $D$, surface tension $\sigma$, and model amplitude
$A_{\mathrm{rep}}$. The resulting trends are summarized in Fig.~\ref{fig:repulsion_vs_capillary_plots}.

Fig.~\ref{fig:repulsion_vs_capillary_plots}(b) reports the ratio
\begin{equation}
R_n \equiv
\frac{\displaystyle \int_{\Omega_{\mathrm{film}}} \bigl| \mathbf{F}_{\mathrm{rep}}\!\cdot\!\mathbf{n} \bigr| \,\mathrm{d}V}
{\displaystyle \int_{\Omega_{\mathrm{film}}} \bigl| \mathbf{F}_{\mathrm{cap}}\!\cdot\!\mathbf{n} \bigr| \,\mathrm{d}V},
\label{eq:RnAbs}
\end{equation}
where $\Omega_{\mathrm{film}}$ denotes the near-contact (thin-film) region identified by activation of the repulsive model,
$\mathbf{n}$ is the local unit normal to the interface (computed from the phase-field gradient), $\mathbf{F}_{\mathrm{rep}}$ is the
repulsive force density (per unit volume) produced by the near-contact interaction, and $\mathbf{F}_{\mathrm{cap}}$ is the capillary
force density. Symbols distinguish the values of $A_{\mathrm{rep}}$, while the
color encodes $\sigma$. The central result is that $R_n$ remains of the same order across the explored parameter space and varies
only weakly with $D$ and with changes in $A_{\mathrm{rep}}$ and $\sigma$, including changes in their ratio $A_{\mathrm{rep}}/\sigma$. This shows that the repulsive action in the film remains consistently
commensurate with the capillary action: the near-contact interaction does not ``over-enforce'' separation when
$A_{\mathrm{rep}}$ is increased, but instead supplies a normal resistance whose magnitude remains comparable to the capillary-driven
normal forcing that promotes interface approach. This is the quantitative signature of a self-tuned near-contact barrier, and it
explains why the overall impact dynamics is largely insensitive to changes in $A_{\mathrm{rep}}$ once the non-coalescing regime is reached.

While panel (b) addresses the relative balance between repulsion and capillarity, panels (c) and (d) clarify how the
absolute level of the repulsive forcing varies when physical parameters are modified. In particular, panel (c) shows the peak
repulsive intensity $\max|\mathbf{F}_{\mathrm{rep}}|$, whereas panel (d) reports the film-averaged magnitude
$\langle|\mathbf{F}_{\mathrm{rep}}|\rangle_{\mathrm{film}}$ (computed as the integral of $|\mathbf{F}_{\mathrm{rep}}|$ over
$\Omega_{\mathrm{film}}$ divided by the activated film volume). When $\sigma$ is decreased, both the peak and the average repulsive
levels decrease accordingly. This is consistent with an adaptive mechanism in which the repulsive forcing scale tracks the capillary
scale: lowering $\sigma$ reduces the capillary forcing level and the near-contact interaction correspondingly relaxes in absolute
amplitude, without requiring a retuning of $A_{\mathrm{rep}}$ to preserve a comparable repulsion--capillarity balance. At the same time,
changes in droplet size mainly affect the geometry of the interaction (how large the active film region becomes) rather than
substantially altering the characteristic repulsive forcing level inside the film, supporting again the interpretation of a regulated barrier.

Finally, panel (e) reports the activated film volume $V_{\mathrm{film}}$ divided by $D^2$. This quantity provides a compact measure
of the extent of the near-contact region, disentangling the trivial size effect (area-like scaling $\sim D^2$) from changes in
activation thickness/patchiness. The weak dependence of $V_{\mathrm{film}}/D^2$ on $D$ indicates that the near-contact region scales
primarily with droplet size, while variations of $\sigma$ and $A_{\mathrm{rep}}$ modulate the degree of activation of the film.
Importantly, the combined view of panels (b) and (e) indicates that the near-contact interaction stabilizes the collision mostly
by adjusting the support of the repulsive action (how many cells participate in $\Omega_{\mathrm{film}}$ at peak activation)
rather than by generating disproportionately large local extremes. This ``support-adjustment'' mechanism provides a natural explanation
for the observed robustness: the near-contact model self-organizes a thin-film repulsive forcing comparable to the capillary forcing,
thereby maintaining a stable, non-coalescing film over a wide range of parameters.

Below a critical $A_{\mathrm{rep}}$, however, the above self-tuning can no longer be maintained: the activated film region becomes intermittent
and collapses, and a topology change (coalescence) may occur even if local repulsive peaks are present. This highlights that stability
requires not only a sufficiently strong local repulsion but also a sufficiently extended and persistent near-contact region at maximum activation.

\subsection{Multi-body interactions: rising bubble swarm in a quiescent environment}
\label{sec:results_multibody}

A central objective of the present near-contact interaction (NCI) framework is to enable \emph{robust many-body interface dynamics} within a diffuse-interface formulation, i.e.\ configurations in which a large number of deformable bubbles undergo repeated close approaches, collisions, and collective rearrangements without triggering spurious topology changes. In this subsection, we therefore demonstrate the capability of the present model to handle dense multi-bubble interactions in three dimensions over long integration times.

To this end, we consider a swarm of bubbles rising in an otherwise quiescent carrier liquid within a fully periodic cubic domain of size
$N_x \times N_y \times N_z = 512^3$ (lattice units). The surface tension is set to $\sigma = 0.01$ (lu), the liquid and gas kinematic viscosities are $\nu_\ell = 10^{-3}$ and $\nu_g = 10^{-2}$ (lu), respectively, and the density ratio is fixed to $\rho_\ell/\rho_g = 1000$.
Buoyancy is modeled through a constant body acceleration $g = 10^{-5}$ (lu/step), applied in the vertical direction.
The initial condition consists of $N_b = 100$ spherical bubbles, corresponding to an overall gas volume occupation of approximately $8\%$.

For this configuration, the dynamical regime can be characterized in terms of dimensionless groups defined using a characteristic bubble diameter $D$ and a characteristic rise velocity $U_b$ (e.g.\ obtained from the mean bubble centroid velocity). In particular, we introduce the bubble Reynolds and Weber numbers
\begin{equation}
Re_b = \frac{U_b D}{\nu_\ell},
\qquad
We_b = \frac{\rho_\ell U_b^2 D}{\sigma},
\label{eq:reb_web_def}
\end{equation}
together with the Galilei number
\begin{equation}
Ga = \frac{\sqrt{g D^3}}{\nu_\ell}.
\label{eq:ga_def}
\end{equation}
In the present simulations, these quantities evaluate to
\begin{equation}
Re_b = 1500,
\qquad
We_b = 37.5 ,
\qquad
Ga   = 1500,
\label{eq:dimless_values_blank}
\end{equation}
where $U_b$ is measured from the mean bubble centroid velocity and $D$ is the initial bubble diameter.

\begin{figure}
    \centering
    \includegraphics[width=0.75\linewidth]{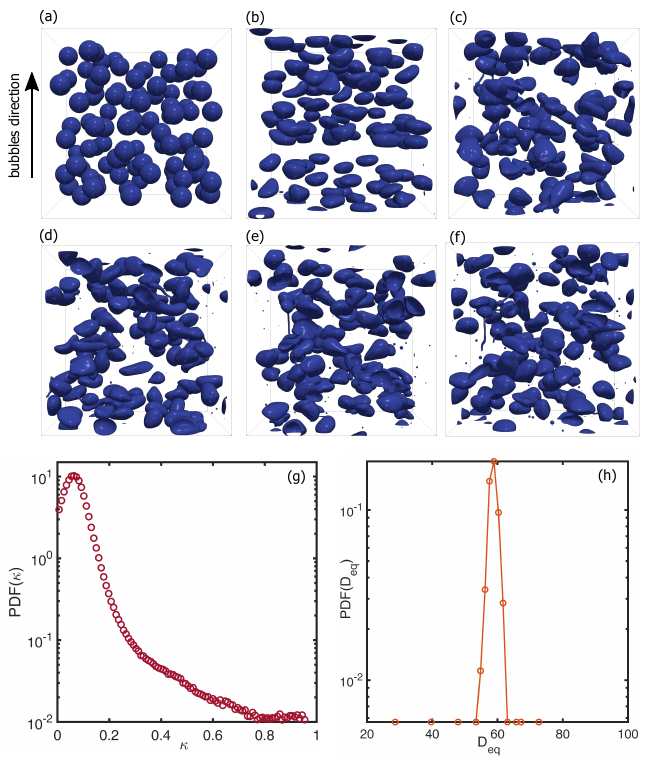}
    \caption{Evolution of a swarm of bubbles rising in a periodic cubic domain. The simulation corresponds to $Re_b = 1500$,  $We_b = 37.5$, and $Ga = 1500$. (a)--(f) Snapshots of the bubble configuration at increasing times. (g) PDF of the surface curvature. (h) PDF of the equivalent diameter. }
    \label{fig:curv_and_rad_pdf}
\end{figure}

Fig.~\ref{fig:curv_and_rad_pdf} provides a qualitative and quantitative overview of the multi-body dynamics. Panels (a)--(f) show six representative snapshots of the bubble field at increasing times. Despite the high number of interfaces and the frequent near-contact events occurring during the collective rise and rearrangement, the bubble population remains well separated, highlighting the effective suppression of coalescence by the short-range repulsion. This is a stringent test for phase-field approaches, as dense swarms can easily develop numerical coalescence when the interfacial transport is not sufficiently robust or when close-range interactions are not properly regularized.

\begin{figure}
    \centering
    \includegraphics[width=0.85\linewidth]{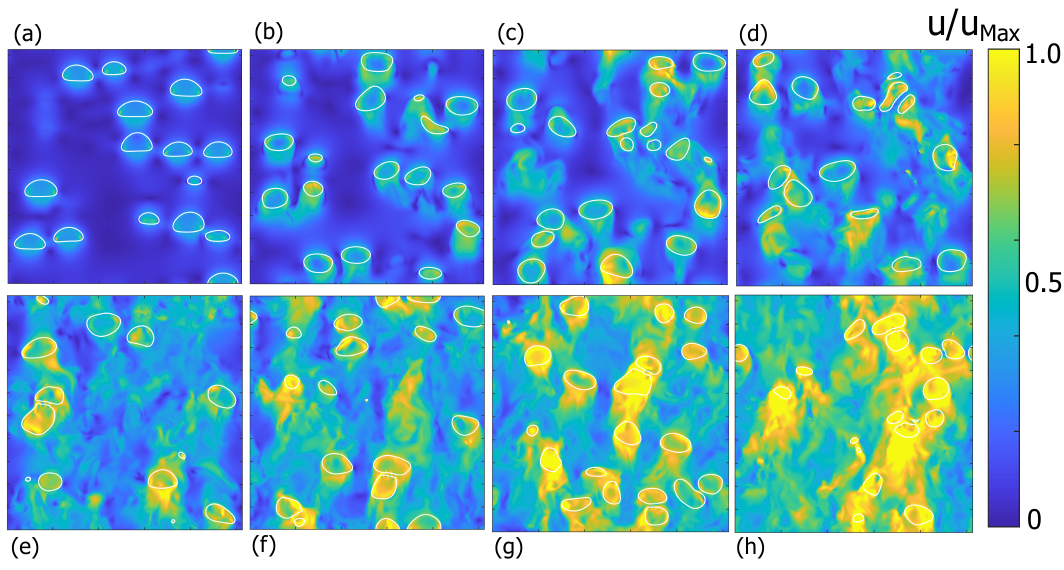}
    \caption{Evolution of the velocity field for the swarm of bubbles in Fig.~\ref{fig:curv_and_rad_pdf}, shown as two-dimensional maps of the normalized velocity magnitude $u/u_{\max}$ on a mid-plane slice at representative times, with bubble contours superimposed. }
    \label{fig:vel_field}
\end{figure}
To quantify the preservation of the dispersed phase, Fig.~\ref{fig:curv_and_rad_pdf} also reports the probability density functions (PDFs) of the interfacial mean curvature $\kappa$ and of the bubble equivalent diameter $D_{\mathrm{eq}}$ (computed from the instantaneous bubble volumes). The curvature distribution exhibits a clear peak associated with the initialization radius $R \simeq 30$ (lu), consistently with the spherical relation for the mean curvature,
\begin{equation}
\kappa = \nabla\cdot\mathbf{n} = \frac{2}{R}
\qquad \Rightarrow \qquad
R = \frac{2}{\kappa},
\label{eq:kappa_radius}
\end{equation}
where $\mathbf{n}$ is the outward unit normal to the interface. Consistently, the diameter PDF is sharply peaked around $D_{\mathrm{eq}} \simeq 2R$, highlighting a quasi-monodisperse population and indicating that, over the time window considered, multi-body interactions predominantly lead to rearrangements and transient deformations rather than coalescence-driven growth. Overall, these statistics confirm that the proposed framework can sustain dense three-dimensional bubble swarms while preserving bubble identity and delivering stable curvature/size distributions suitable for subsequent analyses of dispersity, clustering, and interface-mediated transfer mechanisms.
\begin{figure}
    \centering
    \includegraphics[width=0.5\linewidth]{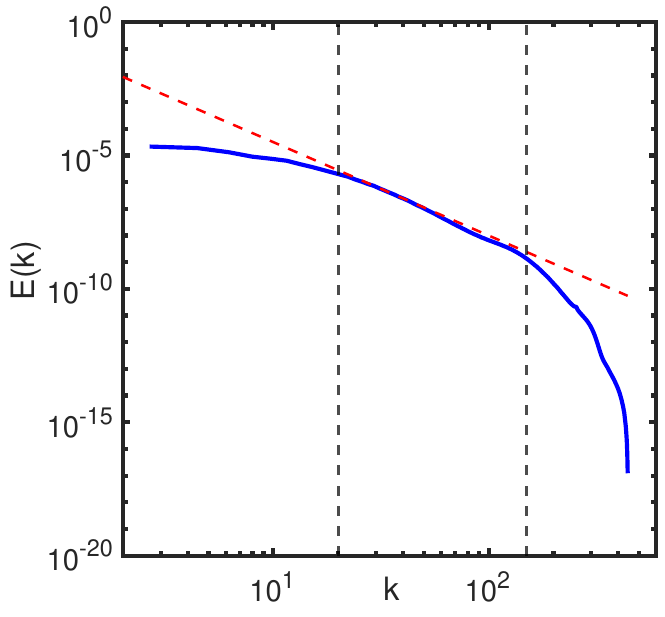}
    \caption{Turbulent kinetic-energy spectrum (blue line) compared with a $k^{-3}$ reference scaling (dashed line).}
    \label{fig:spectrum}
\end{figure}
Complementing the interface-only visualization in Fig.~\ref{fig:curv_and_rad_pdf}, Fig.~\ref{fig:vel_field} reports eight snapshots (panels (a)--(h)) of the carrier-flow response induced by the collective rise of the bubble swarm, with the bubble contours superimposed on a two-dimensional map of the instantaneous velocity field. Starting from an initially quiescent liquid, the motion transitions from a collection of largely isolated bubble wakes to a strongly interacting regime in which wake--wake interactions, repeated near-contact events, and interface deformations generate intense shear layers and spatially intermittent high-velocity patches throughout the domain. As time progresses, these fluctuations become increasingly disordered and space-filling, revealing the onset of a turbulent-like, chaotic flow field sustained by buoyancy-driven deformable bubbles (often referred to as \emph{pseudo-turbulence}~\cite{almeras2017experimental,van1998pseudo,mercado2010bubble,hosokawa2013bubble}).  
To provide a first quantitative indicator of the turbulence-like agitation suggested by the flow snapshots in Fig.~\ref{fig:vel_field}, Fig.~\ref{fig:spectrum} reports the isotropic spectrum of turbulent kinetic energy $E(k)$ of the carrier-phase velocity fluctuations, compared with a $k^{-3}$ reference scaling (dashed line). Over an intermediate band of wavenumbers (delimited by the vertical dashed lines), the spectrum develops a clear power-law decay close to $k^{-3}$, a hallmark scaling frequently reported in bubble-driven pseudo-turbulence and commonly associated with wake-dominated dynamics at scales at and below the bubble size (i.e. for $k \gtrsim k_b\sim 2\pi/d_b$). At the largest scales (small $k$), deviations from the reference slope reflect the finite domain size and the anisotropic nature of buoyancy-driven forcing, whereas at high wavenumbers the spectrum steepens due to viscous dissipation and finite-resolution effects.

A dedicated quantitative characterization of the resulting pseudo-turbulent regime, including anisotropy, intermittency, and scale-by-scale budgets, is deferred to future work.

\section{Conclusions}

In this work, we have introduced a near-contact interaction (NCI) model for conservative Allen--Cahn phase-field lattice Boltzmann simulations of multiphase flows. The central objective was to provide a physically grounded and computationally efficient mechanism to prevent spurious coalescence in situations where the outcome of interface interaction is controlled by an unresolved thin intervening film. The proposed formulation augments the momentum equation with a short-range repulsive contribution that activates only when two diffuse interfaces approach each other with opposite orientations, thereby mimicking the stabilizing action of unresolved thin-film physics within a mesoscopic diffuse-interface framework. 

A distinctive feature of the present approach is that the film thickness is not obtained by ray tracing or geometric reconstruction along interface normals. Instead, an effective local film-thickness indicator is derived in closed form from the phase-field overlap using the equilibrium hyperbolic-tangent profile. This construction preserves locality, improves symmetry of the interaction, and is particularly well-suited for massively parallel GPU-oriented implementations. In combination with the thread-safe recursive-regularized lattice Boltzmann solver and the conservative Allen--Cahn transport equation, the resulting framework remains fully local, scalable, and robust even in the presence of strong density and viscosity contrasts. 

The numerical results show that the model captures the correct non-coalescing outcome in droplet-collision benchmarks, including both head-on and off-axis impacts, with excellent agreement with experimental observations. Beyond reproducing the interface kinematics, the simulations recover the expected thin-film hydrodynamic signatures associated with rebound, including suppression and inversion of the film-normal flow and the emergence of stagnation-like regions near the interaction zone. These results indicate that the NCI does not merely act as an artificial penalty, but reproduces the hydrodynamic conditions required for sustained interfacial separation and bounce. 

An additional important outcome is the evidence of a self-tuning behavior of the repulsive interaction. Once the repulsive amplitude exceeds the threshold required to avoid coalescence, further increases produce little change in the macroscopic collision dynamics. The analysis of force balances within the activated film region shows that the repulsive contribution remains of the same order as the capillary forcing, while the model adapts predominantly through the spatial support of the activated near-contact region rather than through unbounded local force amplification. This property is particularly appealing because it reduces the sensitivity of the macroscopic dynamics to parameter tuning and suggests that the model behaves as an adaptive near-contact barrier rather than as a stiff numerical constraint. 

Finally, the rising-bubble-swarm simulations demonstrate that the method extends naturally to dense many-body configurations in three dimensions, where repeated close approaches and collective rearrangements occur over long integration times. In this regime, the framework preserves bubble identity, avoids spurious topology changes, and sustains complex pseudo-turbulent carrier-phase agitation, thereby opening the way to systematic studies of dense bubbly flows, emulsions, and other strongly interacting disperse systems. 

Overall, the present results indicate that the proposed LB--Allen--Cahn--NCI framework provides an effective compromise between physical fidelity, numerical robustness, and computational efficiency for unresolved near-contact dynamics in diffuse-interface multiphase simulations. Future work will focus on a more quantitative calibration of the repulsive closure against thin-film drainage theory and experiments, on extensions to wetting and wall-mediated near-contact events, and on the application of the method to turbulent many-body flows where collisions, rebounds, and topology preservation play a central role in the resulting complex dynamics.

\section*{Acknowledgements}

A.M. and M.L. acknowledge funding from the Italian Government through the PRIN (Progetti di Rilevante Interesse Nazionale) Grant (MOBIOS), ID: 2022N4ZNH3, CUP: F53C24001000006, and computational support from CINECA through the ISCRA B project MIPLAST (IsB31, HP10BZY7BK). S.S. acknowledges the support from the European Research Council under the ERCPoC Grant No. 101187935 (LBFAST).
M.L. acknowledges the support of the Italian National Group for Mathematical Physics (GNFM-INdAM).

\section*{Conflict of interest}
The authors report no conflict of interest.


\end{document}